\documentclass{article}

\usepackage{microtype}
\usepackage{graphicx}
\usepackage{subfigure}
\usepackage{booktabs} %
\usepackage{amsmath}
\usepackage{amssymb}
\usepackage{mathtools}
\usepackage{subcaption}
\usepackage{array}
\usepackage{amsthm}
\usepackage{microtype}
\usepackage{subfigure}
\usepackage{color}
\usepackage{url}
\usepackage{wrapfig}
\usepackage{booktabs}       %
\usepackage{nicefrac}       %
\usepackage{microtype}      %
\usepackage{xcolor}         %
\usepackage{amsmath,amssymb,amsfonts}
\usepackage{multirow}
\usepackage{microtype}
\usepackage{wrapfig}
\usepackage{graphicx}
\usepackage{caption}
\usepackage{booktabs}
\usepackage{bm}
\usepackage{enumitem}
\usepackage{cases}
\usepackage{subcaption}
\usepackage{algorithm}
\usepackage{pifont}
\usepackage{booktabs} %
\usepackage{hyperref}

\usepackage[accepted]{icml2024}

\usepackage{amsmath}
\usepackage{amssymb}
\usepackage{mathtools}
\usepackage{amsthm}

\usepackage[capitalize,noabbrev]{cleveref}

\theoremstyle{plain}
\newtheorem{theorem}{Theorem}[section]

\theoremstyle{definition}

\theoremstyle{remark}

\usepackage[textsize=tiny]{todonotes}

\icmltitlerunning{TERD: A Unified Framework for Safeguarding Diffusion Models Against Backdoors}

\begin{document}

\twocolumn[
\icmltitle{TERD: A Unified Framework for Safeguarding Diffusion Models \\ Against Backdoors}

\icmlsetsymbol{equal}{*}

\begin{icmlauthorlist}
\icmlauthor{Yichuan Mo}{yyy}
\icmlauthor{Hui Huang}{sch}
\icmlauthor{Mingjie Li}{mj}
\icmlauthor{Ang Li}{sch}
\icmlauthor{Yisen Wang}{yyy,ai}
\end{icmlauthorlist}

\icmlaffiliation{yyy}{National Key Lab of General Artificial Intelligence, School of Intelligence Science and Technology, Peking University, China}
\icmlaffiliation{ai}{Institute for Artificial Intelligence, Peking University, China}
\icmlaffiliation{mj}{CISPA Helmholtz Center for Information Security, Germany}
\icmlaffiliation{sch}{School of EECS, Peking University, China}
\icmlcorrespondingauthor{Yisen Wang}{yisen.wang@pku.edu.cn}

\icmlkeywords{Machine Learning, ICML}

\vskip 0.3in
]

\printAffiliationsAndNotice{}  %

\begin{abstract}
Diffusion models have achieved notable success in image generation, but they remain highly vulnerable to backdoor attacks, which compromise their integrity by producing specific undesirable outputs when presented with a pre-defined trigger. In this paper, we investigate how to protect diffusion models from this dangerous threat. Specifically, we propose \textbf{TERD}, a backdoor defense framework that builds unified modeling for current attacks, which enables us to derive an accessible reversed loss. A trigger reversion strategy is further employed: an initial approximation of the trigger through noise sampled from a prior distribution, followed by
refinement through differential multi-step samplers. Additionally, with the reversed trigger, we propose backdoor detection from the noise space, introducing the first backdoor input detection approach for diffusion models and a novel model detection algorithm that calculates the KL divergence between reversed and benign distributions. Extensive evaluations demonstrate that TERD secures a 100\% True Positive Rate (TPR) and True Negative Rate (TNR) across datasets of varying resolutions. TERD also demonstrates nice adaptability to other Stochastic Differential Equation (SDE)-based models. Our code is available at \url{https://github.com/PKU-ML/TERD}.
\end{abstract}

\section{Introduction}
In recent years, we have witnessed significant advancements in generative models \cite{goodfellow2014generative,kingma2013auto,kingma2018glow}, with diffusion models emerging as a particularly notable representative \cite{ho2020denoising,rombach2021highresolution,yang2023diffusion}. These models have demonstrated their marvelous performances in a diverse range of applications from image generation \cite{rombach2021highresolution}, content editing \cite{meng2022sdedit}, zero-shot classification \cite{li2023your} to adversarial purification \cite{nie2022diffusion}. However, the widespread application of diffusion models raises concerns about their security issues like backdoor attacks \cite{gu2017badnets,chen2017targeted}, where models can be manipulated to produce harmful outputs under specific conditions, posing significant legal and ethical risks. 
Therefore, in this paper, we explore how to defend against backdoor attacks for diffusion models, which is less investigated before.

Unlike common classification models, diffusion models operate on noise outputs rather than class logits, making them impervious to conventional defenses \cite{wu2021adversarial,wu2022backdoorbench} designed for classification tasks. The challenge is exacerbated by the complexity of their input-output dynamics over various timesteps, e.g., the model's behavior changes across different timesteps and the underlying formulation is often inaccessible to defenders. This significantly hinders the ability to effectively identify and mitigate backdoor triggers.

To address this challenge, in this paper, we propose a novel defense strategy that begins by systematically characterizing existing backdoor attacks in diffusion models. Our approach involves creating unified formulations of backdoor attacks, enabling us to derive an accessible reversed loss. For the accessibility of inputs, we introduce a two-stage trigger reversion process: we first estimate the trigger using noise sampled from a prior distribution, followed by refinement through differential multi-step samplers. This process allows for accurate identification and neutralization of backdoor inputs. Equipped with the estimated trigger, we can detect backdoor attacks from both the input and model perspectives in the noise space, leveraging the statistical characteristics of noise distributions to distinguish between benign and malicious inputs. We refer to this comprehensive defense framework as TERD (Trigger Estimation and Refinement for Diffusion). TERD has demonstrated remarkable success across diverse datasets, achieving a 100\% True Positive Rate (TPR) and True Negative Rate (TNR). Further, TERD works well against a wide range of attack scenarios, including those with varied poisoning rates, trigger sizes, and even sophisticated adaptive attacks. Beyond the diffusion models, TERD also shows promise for defending other Stochastic Differential Equation (SDE)-based models against backdoor attacks. In summary, our main contributions are listed as follows:
\vspace{-6pt}
\begin{itemize}
    \item We specially design a novel trigger reversion algorithm based on the unified modeling against backdoor attacks in diffusion models, which can accurately reverse triggers with high quality.
    \vspace{-10pt}
    \item With the reversed trigger, we develop an input and model detection method in the noise space to protect the diffusion models from backdoors.
    \vspace{-10pt}
    \item Extensive experiments show the efficacy of our defense across varied scenarios and its potential applicability to broader SDE-based generative models.
\end{itemize}

\section{Related Work}
\subsection{Backdoor Attacks in Diffusion Models}
Backdoor attacks, also known as Trojan attacks \cite{gu2017badnets, chen2017targeted}, were initially studied in the context of classification models. These attacks involve implanting pre-defined malicious behaviors into neural networks. While the victim models maintain normal functionality with benign inputs, the presence of a trigger in the input causes the model to exhibit malicious behaviors, such as misclassification or illegal content generation.
Recent studies, such as \citet{chou2023backdoor} and \citet{chen2023trojdiff}, have demonstrated that diffusion models are also vulnerable to these attacks. In these scenarios, a trigger is added to noise sampled from a prior distribution. Images generated from this altered noise become target images, resulting in unexpected sequences. VillianDiffusion \cite{chou2023villandiffusion} further extends it to continuous diffusion models.
Additional research has shown that backdoor attacks can be executed using natural language prompts \cite{zhai2023text, huang2023zero, struppek2023rickrolling} (specifically for text-to-image diffusion models) or by poisoning the training set \cite{pan2023trojan}. However, these attacks can be easily defended by purifying the text encoder or additional human inspection. Therefore, in this paper, we focus on defending against backdoor attacks from the pixel level, which not only has good stealthiness but also endangers all existing diffusion models.

\subsection{Existing Backdoor Defense}
\label{sec:existing_defense}
Similar to defenses against adversarial attacks \cite{li2020implicit,wang2019dynamic,wang2020improving,wu2020adversarial,mo2022adversarial}, current backdoor defenses mainly focus on classification models. These defenses can be categorized into two types: input-level and model-level defenses. Input-level defenses aim to detect whether an input sample is a backdoor sample. Previous studies have shown that backdoor samples can be identified through neural activations \cite{chen2018detecting} or frequency analysis \cite{zeng2021rethinking}. Techniques from other fields, such as differential privacy and explainable visualization tools, further enhance detection success rates \cite{doan2020februus, du2019robust}, as backdoor samples often appear as outliers relying on local spurious features. Model-level defenses work by first detecting whether a model has been implanted with a backdoor and then mitigating the backdoor effect. Regarding backdoors as shortcuts between the real and target classes, methods like \cite{wang2019neural,tao2022better,hu2021trigger} employ reverse engineering by maximizing the classification loss across all classes to identify potential triggers. 
Once the model is identified as backdoored, purification-based defenses such as fine-tuning \cite{sha2022fine,xiong2023rethinking}, pruning \cite{wu2021adversarial,chai2022one}, or unlearning \cite{liu2022backdoor,wei2023shared} are employed to reduce the attack success rate while maintaining benign accuracy. However, these defenses fail to protect diffusion models because the input to a diffusion model is Gaussian noise rather than natural images, and diffusion models predict added Gaussian noise rather than discriminative results of natural images. 

The most relevant work to ours is Elijah \cite{an2023remove}, the method designed specifically for backdoor defense in diffusion models. However, Elijah does not establish a unified loss for current attacks, assuming the trigger is part of the model output, which does not apply to state-of-the-art attacks such as TrojDiff \cite{chen2023trojdiff}. Additionally, Elijah's model detection method assumes that backdoor models generate images with higher similarity, a claim contradicted by \citet{chen2023trojdiff}, which demonstrates that target images can consist of multiple images with diverse and colorful patterns.

\section{Preliminary}
\label{sec:diff_Prcs}

\subsection{Discrete Diffusion Model}
\label{sec:dis}

Based on the Markov chain, Denoising Diffusion Probabilistic Models (DDPM) \cite{ho2020denoising} connects the data and prior distribution (e.g., Gaussian distribution) by defining a forward diffusion and backward denoising process. In its forward process, Gaussian noise is gradually added to images and the conditional distribution $p(\mathbf{x}_t|\mathbf{x}_{t-1})$ is defined as $\mathcal{N}(\sqrt{\alpha_t}\mathbf{x}_{t-1},(1-\alpha_t)\mathbf{I})$ where $\alpha_t\in(0,1)$. According to Bayes Rule, given $\mathbf{x}_0$, we can sample $\mathbf{x}_t$ of timestep $t$ (0$<t\leq T$) directly from the following equation:
\begin{equation}
\label{eq:ddpm_forard}
\mathbf{x}_t=\sqrt{\Bar{\alpha}_t}\mathbf{x}_{0}+\sqrt{1-\Bar{\alpha}_t}\bm{\epsilon}, \quad\bm{\epsilon}\sim\mathcal{N}(0,\mathbf{I}),
\vspace{-5pt}
\end{equation}
where $\Bar{\alpha}_t=\prod\limits_{i=1}^{t}\alpha_i$. The boundary conditions require that $\lim\limits_{t\rightarrow T}\Bar{\alpha}_t= 0$ to ensure that $p(\mathbf{x}_t|\mathbf{x}_{0})$ converges to $\mathcal{N}(0,\mathbf{I})$. Therefore, in the denoising process, we first sample $\mathbf{x}_T$ from $\mathcal{N}(0,\mathbf{I})$ and then generate $\mathbf{x}_{t-1}$ step-by-step using $p(\mathbf{x}_{t-1}|\mathbf{x}_{t},\mathbf{x}_{0})=\frac{p(\mathbf{x}_t|\mathbf{x}_{t-1},\mathbf{x}_0)p(\mathbf{x}_{t-1}|\mathbf{x}_0)}{p(\mathbf{x}_t|\mathbf{x}_0)}$. According to Equation \ref{eq:ddpm_forard}, we can estimate $\mathbf{x}_0$ with $\frac{\mathbf{x}_t-\sqrt{1-\Bar{\alpha}_t}F_\theta(\mathbf{x}_t,t)}{\sqrt{\Bar{\alpha}_t}}$ once the network $F_\theta$ predicts $\bm{\epsilon}$:
\begin{equation}
    \min_{\theta}\vert\vert F_\theta(\mathbf{x}_{t},t)-\bm{\epsilon}\vert\vert_2.
\end{equation}

\subsection{Continuous Diffusion Model}
\label{sec:con}
In \citet{song2020score}, a unified Stochastic Diffusion Equation (SDE)-based framework is proposed to encapsulate the diffusion model. When $t$ becomes continuous, the diffusion process is characterized by the following forward SDE: 
\begin{equation}
    d\mathbf{x}_t=\mathbf{f}(\mathbf{x}_t,t)dt+g(t)d\mathbf{w}, 
\end{equation}
where $t\in[0,T]$ and $\mathbf{f}(\mathbf{x}_t,t)$, $g(t)$ are the drift and diffusion coefficients, respectively. According to \citet{anderson1982reverse}, the denoising process corresponds to a reversed SDE:
\begin{equation}
\label{eq:rev_sde}
    d\mathbf{x}_t=[\mathbf{f}(\mathbf{x}_t,t)-g(t)^2\nabla_\mathbf{x}\log p_{t}(\mathbf{x}_t)]dt+g(t)d\mathbf{w}.
\end{equation}
We cannot solve the above equation directly due to the existence of term $\nabla_\mathbf{x}\log p_{t}(\mathbf{x}_t,\bm{\epsilon})$. However, in the forward diffusion process, we can train the model $F_\theta$ with $\mathbf{x}_t$ and the time step $t$ to fit it:
\begin{equation}
    \min_{\theta}\vert\vert F_\theta(\mathbf{x}_{t},t)-\nabla_\mathbf{x}\log p_{t}(\mathbf{x}_t)\vert\vert_2.
\end{equation}
Thus in the sampling stage, we can generate images by solving Equation \ref{eq:rev_sde} with appropriate samplers, such as Heun solver \cite{karras2022elucidating} and DPM solver \cite{lu2022dpm}.

\begin{table*}[t]
    \centering
    \caption{Designed choices adopted by current attacks and their relationship to our unified formulation. As long as the coefficient cannot be derived from the benign diffusion process in one of the attacks, we consider it inaccessible to the defenders.}
    \resizebox{0.9\linewidth}{!}{
    \begin{tabular}{cc|ccc|cccccc}
    \toprule
    & &BadDiffusion  & TrojDiff & VillanDiffusion  &  \multirow{2}*{Accessible to defenders}\\
    & &\cite{chou2023backdoor} & \cite{chen2023trojdiff} & \cite{chou2023villandiffusion} \\
    \midrule
    \multirow{3}*{Diffusion Process}&$a(\mathbf{x}_0,t)$&$\sqrt{\Bar{\alpha}_t}$ &  $\sqrt{\Bar{\alpha}_t}$&$\int_0^t\mathbf{f}(\mathbf{x}_t,t)dt/\mathbf{x}_0+1$& \checkmark  \\
    & $b(t)$&$\sqrt{1-\Bar{\alpha}_t}$ &$\sqrt{1-\Bar{\alpha}_t}$&$\sqrt{\int_0^tg^2(t)dt}$ &\checkmark\\
    & $c(t)$&$1-\sqrt{\Bar{\alpha}_t}$ &$\sqrt{1-\Bar{\alpha}_t}$&$\int_0^t H(t)dt$ &\ding{55}\\
    \midrule
        \multirow{2}*{Training Loss}&$f(\textbf{x}_t,\epsilon)$&$\bm{\epsilon}$ & $\bm{\epsilon}$ & $\nabla_\mathbf{x}\log p_{t}(\mathbf{x}_t,\bm{\epsilon})$ &   \checkmark\\
    & $d(t)$&$\frac{\sqrt{1-\Bar{\alpha}_t}}{1+\sqrt{\alpha_t}}$ &0&$\frac{H(t)}{g(t)^2}$ &\ding{55}\\
    \toprule
    \end{tabular}}
    \label{tab:unified_forward}
\end{table*}

\subsection{Backdoor Diffusion Model}
\label{sec:bd_diff}

Only a few works, such as \citet{chou2023backdoor,chen2023trojdiff,chou2023villandiffusion}, explored backdoor attacks in diffusion models. In their threat models, attackers have access to the training process of diffusion models. They develop a backdoor diffusion process to ensure that when a trigger is attached to the sampled noise, the generated images transform into predefined target images. The trigger and target images are tensors with the same shape as benign images and are inaccessible to defenders. To maintain the benign utility of the model, the benign training loss, as defined in Sections \ref{sec:dis} and \ref{sec:con}, is also incorporated into the training process.

\textbf{BadDiffusion} \cite{chou2023backdoor}. Designed for discrete diffusion models, BadDiffusion inserts backdoors by gradually attaching triggers to noisy images. Its backdoor diffusion process is defined as:
\begin{equation}
\mathbf{x}_t=\sqrt{\bar{\alpha}}_t \mathbf{x}_0+\sqrt{1-\bar{\alpha}_t}\bm{\epsilon}+(1-\sqrt{\bar{\alpha}_t})\mathbf{r},
\end{equation}
where $\mathbf{x}_0$ refers to target images instead of benign images, and $\mathbf{r}$ is the trigger. 

\textbf{TrojDiff} \cite{chen2023trojdiff}. Similar to BadDiffusion, TrojDiff aims to insert backdoors into discrete diffusion models. However, it introduces both patch-based and whole-image triggers using a new variable, $\bm{\gamma}$. The backdoor diffusion process of TrojDiff is formulated as:
\begin{equation}
\mathbf{x}_t=\sqrt{\bar{\alpha}_t}\mathbf{x}_0+\sqrt{1-\bar{\alpha}_t}\bm{\gamma\epsilon}+\sqrt{1-\bar{\alpha}_t}\mathbf{r}.
\end{equation}
\textbf{VillanDiffusion} \cite{chou2023villandiffusion}. VillanDiffusion develops a backdoor attack for continuous diffusion models. The backdoor SDE is modified from the benign forward SDE to incorporate the trigger into the backdoor diffusion process: 
\begin{equation}
d\mathbf{x}_t=\mathbf{f}(\mathbf{x}_t,t)dt+H(t)\mathbf{r}+g(t)d\mathbf{w},
\end{equation}
where $H(t)$ is a continuous function inaccessible to defenders and meets the boundary condition $\int_0^t H(t)dt=1$ to ensure the backdoor attack can be accurately triggered by $\mathbf{r}$.

\section{Reverse Engineering}

\subsection{A Unified Loss for Trigger Reversion}
\label{sec:unified}
As summarized in Section \ref{sec:diff_Prcs}, in addition to the benign diffusion process, current backdoor attacks for diffusion models define an additional diffusion process \textit{i.e.}, backdoor diffusion process for target image generation. Despite the differences in the details among the attacks, we can unify their formulations with the following equation\footnote{The blending coefficient $\gamma$ is omitted for TrojDiff because we regard it as part of the trigger and optimize it for TrojDiff during the trigger reversion process.}:
\begin{equation}
\label{eq:t_input}
    \mathbf{x}_t = a(\mathbf{x}_0,t)\mathbf{x}_0+b(t)\bm{\epsilon}+c(t) \mathbf{r}.
\end{equation}
Here, $a(\mathbf{x}_0,t)$ and $b(t)$ are two coefficients that follow the benign diffusion process and the backdoor coefficient $c(t)$ is defined by attackers. To ensure that the backdoor effect can be triggered by $\mathbf{r}$, $c(t)$ needs to first satisfy the following boundary condition: $\lim\limits_{t\rightarrow T}{c}(t)=1$. In addition, with the initial condition: $\mathbf{x}_t=\mathbf{x}_0$, we can get: $\lim\limits_{t\rightarrow 0}c(t)=0$. According to the formulations in Section \ref{sec:bd_diff}, we summarize their corresponding relations with existing attacks in Table \ref{tab:unified_forward}. Meanwhile, we also established a unified form of backdoor training loss for those attacks:
\begin{equation}
\label{eq:all_loss}
\min_{\theta}\mathbb{E}_{t,\bm{\epsilon}}\vert\vert F_\theta(\mathbf{x}_{t},t)-f(\mathbf{x}_t,\bm{\epsilon})+d(t)\mathbf{r}\vert\vert_2,
\end{equation}
where $f(\mathbf{x}_t,\bm{\epsilon})$ is the training target for the benign loss. For example, for the DDPM model, it denotes the gaussian noise added to the noisy image. The detailed formulation of $d(t)$ is related to the specific attack adopted by attackers, such as $d(t)\equiv0$ for TrojDiff and a black-box function for VillanDiffusion.  Therefore, it indicates that it is not feasible to reverse the trigger directly through Equation \ref{eq:all_loss}. Note that in Elijah \cite{an2023remove}, they heuristically assume $d(t)=0.5$ and make a trade-off between BadDiffusion and TrojDiff ($\lim\limits_{t\rightarrow T}\frac{\sqrt{1-\Bar{\alpha}_t}}{1+\sqrt{\alpha_t}}=1$ for BadDiffusion). This could lead to the failure of defense, particularly in some difficult cases. Therefore, it is necessary to first establish a unified loss to more accurately characterize the relation between the trigger and the model output. Observe that for Equation \ref{eq:all_loss}, we can divide it with the losses of two independent noises $\bm{\epsilon}_1$, $\bm{\epsilon}_2$ respectively. Furthermore, we can employ the triangle inequality to obtain a lower bound for direct optimization:
\begin{equation}
\label{eq:elbo}
\begin{aligned}   \mathbb{E}_{t,\bm{\epsilon}_1,\bm{\epsilon}_2}\frac{1}{2}\vert\vert F_\theta(\mathbf{x}_{t}&(\bm{\epsilon}_1, \mathbf{r}),t)-f(\mathbf{x}_t(\bm{\epsilon}_1, \mathbf{r}),\bm{\epsilon}_1)+d(t)\mathbf{r}\vert\vert_2 \\
    +  \frac{1}{2}\vert\vert F_\theta(\mathbf{x}_{t}&(\bm{\epsilon}_2, \mathbf{r}),t)-f(\mathbf{x}_t(\bm{\epsilon}_2, \mathbf{r}),\bm{\epsilon}_2)+d(t)\mathbf{r}\vert\vert_2 \\
      \geq \frac{1}{2}\mathbb{E}_{t,\bm{\epsilon}_1,\bm{\epsilon}_2} &\vert\vert F_\theta(\mathbf{x}_{t}(\bm{\epsilon}_1, \mathbf{r}),t)-f(\mathbf{x}_t(\bm{\epsilon}_1, \mathbf{r}),\bm{\epsilon}_1)\\  -F_\theta(\mathbf{x}_{t}&(\bm{\epsilon}_2, \mathbf{r}),t)+f(\mathbf{x}_t(\bm{\epsilon}_2, \mathbf{r}),\bm{\epsilon}_2)\vert\vert_2.
\end{aligned}
\end{equation}
Due to the non-negative property of the norm operation, when Equation \ref{eq:all_loss} is optimized to 0, the lower bound in Equation \ref{eq:elbo} also reaches a minimum point. It means that we can substitute Equation \ref{eq:all_loss} with Equation \ref{eq:elbo} for trigger reversion. To avoid $r$ collapses to the full-zero vector, we introduce ${l}_1$ norm for penalization and $\lambda$ as the trade-off coefficient:
\begin{equation}
\label{eq:loss_final}
\small
\begin{aligned}
&\mathcal{L}(\mathbf{r}, \mathbf{x}_{t}) =   \vert\vert (F_\theta(\mathbf{x}_{t}(\bm{\epsilon}_1, \mathbf{r}),t)-f(\mathbf{x}_t(\bm{\epsilon}_1, \mathbf{r}),t),\bm{\epsilon}_1)\\& -F_\theta(\mathbf{x}_{t}(\bm{\epsilon}_2, \mathbf{r}),t)+f(\mathbf{x}_t(\bm{\epsilon}_2, \mathbf{r}),t),\bm{\epsilon}_2)\vert\vert_2-\lambda\vert\vert \mathbf{r}\vert\vert_2 
\end{aligned}
\end{equation}
Note that Equation \ref{eq:loss_final} unifies the expression of all current attacks from the reversed loss, free of the trade-off between the detailed formulations. In order to obtain a high-quality reversed trigger, our proposed reverse engineering approach is composed of the following two steps, including the preliminary estimation of the trigger with a surrogate distribution and further refinement with a differential generation process.

\subsection{Trigger Estimation}

\begin{figure*}[t]

    \centering
     \resizebox{1.0\linewidth}{!}{
	\includegraphics[width=0.23\linewidth]{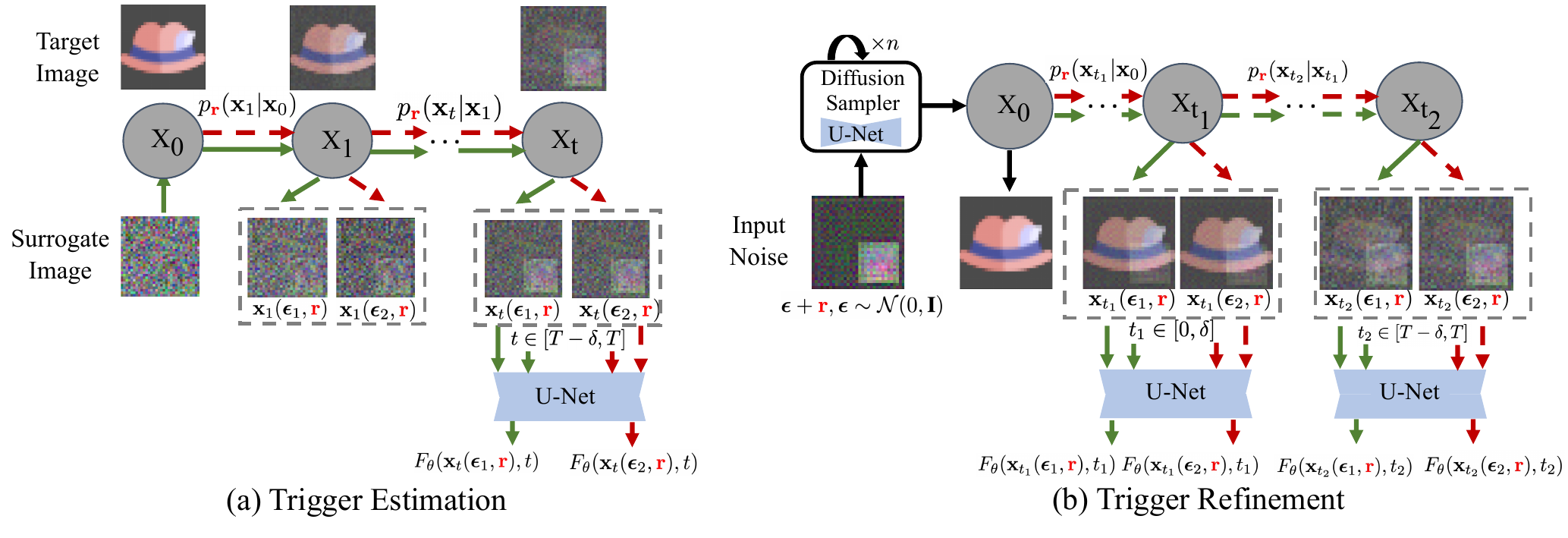}
}
 \vspace{-17pt}
    \caption{An illustration for our proposed reverse engineering method.}
        \vspace{-10pt}
\label{fig:long_ft}
\end{figure*}

Although in Equation \ref{eq:loss_final}, we built a unified loss to eliminate the difference in formulations for various attacks, it still needs further improvement to perform reverse engineering. The obstacle is that $\mathbf{x}_t$ is unknown to defenders, which is simultaneously decided by the target images $\mathbf{x}_0$ and the coefficient $c(t)$. However, the property of diffusion models guarantees that when $t$ approaches $T$, $\mathbf{x}_t$ will converge to the prior distribution that is little affected by $\mathbf{x}_0$. Therefore, we can substitute $\mathbf{x}_0$ with a surrogate image $\hat{\mathbf{x}}_0$ sampled from a substitute distribution, \textit{e.g.}, the standard gaussian distribution $\hat{p}_{prior}$, to estimate $\mathbf{x}_t$. Here, we also prove this property from a theoretical perspective:
\begin{theorem}
\label{theory:1}
Given the target image $\mathbf{x}_0\sim p_{target}$ and a surrogate image $\hat{\mathbf{x}}_0\sim\hat{p}_{prior}$, let $\mathbf{p}_t$ and $\mathbf{q}_t$ denotes the distribution of $\mathbf{x}_0$ and $\hat{\mathbf{x}}_0$ at timestep $t$. we can prove that:
\vspace{-2pt}
\begin{equation}
\label{eq:theory}
    \frac{\partial D_{KL}(\mathbf{p}_t||\mathbf{q}_t)}{\partial t}\leq0.
\vspace{-8pt}
\end{equation}
\end{theorem}
For the proof of Theorem \ref{theory:1}, please refer to Appendix \ref{ap:prof} for details. Following \cite{song2021maximum,nie2022diffusion}, we first prove that the current backdoor diffusion processes are all Wiener Processes. Then we finished the proof with its property. Equation \ref{eq:theory} means that the divergence between $\mathbf{p}_t$ and $\mathbf{q}_t$ will monotonically decrease with $t$ during the diffusion process. Thus $\mathbf{p}_t$ and $\mathbf{q}_t$ will become indistinguishable when $t$ is large. Therefore, for $t\in[T-\delta,T]$ and $\delta\ll T$, we can substitute $\mathbf{x}_0$ with $\hat{\mathbf{x}}_0$ and simplify Equation \ref{eq:t_input} to the following equation:
\vspace{-3pt}
\begin{equation}
\label{eq:pi_for}
 \mathbf{x}^{(1)}_t =a(\mathbf{\hat{x}}_0,t)\mathbf{\hat{x}}_0+b(t)\bm{\epsilon}+\mathbf{r}. 
 \vspace{-3pt}
\end{equation}
Here we omit $c(t)$ because $c(t)\approx1$ when $t\in[T-\delta,T]$. Substituting $\mathbf{x}_t$ in Equation \ref{eq:loss_final} with $\mathbf{x}^{(1)}_t$, and we can get:
\vspace{-3pt}
\begin{equation}
\small
\label{eq:loss_pi}
\begin{aligned}
    &\mathcal{L}_1(\mathbf{r}) = \vert\vert F_\theta(\mathbf{x}_{t}^{(1)}(\bm{\epsilon}_1,\mathbf{r}),t)-f(\mathbf{x}^{(1)}_t(\bm{\epsilon}_1,\mathbf{r}),t),\bm{\epsilon}_1)\\ & -F_\theta(\mathbf{x}_{t}^{(1)} (\bm{\epsilon}_2,\mathbf{r}),t)+f(\mathbf{x}^{(1)}_t(\bm{\epsilon}_2,\mathbf{r}),t),\bm{\epsilon}_2)\vert\vert_2-\lambda\vert\vert \mathbf{r}\vert\vert_2.
\vspace{-6pt}
\end{aligned}
\end{equation}
Directly optimizing it with a commonly used optimizer such as SGD \cite{bottou2010large}, we can preliminarily reverse the trigger. However, if we could represent $\mathbf{x}_0$ with a more precise formulation, the quality of the reversed trigger could be further improved.

\subsection{Trigger Refinement}

Recall that in those early studies for diffusion models, the sampling processes are time-consuming because they follow the reversed Markovian chain, which consists of thousands of steps. To save the computational cost, following-up works, such as the Denoising Diffusion Implicit Model (DDIM) sampler \cite{song2020denoising} propose that multiple denoised steps are equal to a non-Markovian process with fewer steps. It indicates that we can obtain high-quality images even with a few steps of sampling. Note that the operations in the denoised process are all differential. Thus it motivates us to estimate $\mathbf{x}_t$ with multi-step generations. If $\Phi_n(\cdot)$ denotes n-steps DDIM sampler \footnote{For continuous diffusion models, it denotes $n$ steps Heun sampler}, we can obtain the target image $\mathbf{x}_0$ with the trigger $\mathbf{r}$:
\vspace{-3pt}
\begin{equation}
    \mathbf{x}_0 = \Phi_n(\mathbf{r})
\vspace{-3pt}
\end{equation}
Similar to Equation \ref{eq:pi_for}, we can obtain a more precise formula for $\mathbf{x}_t$ when $t\in[T-\delta,T]$ and $\delta\ll T$:
\vspace{-3pt}
\begin{equation}
\mathbf{x}^{(2)}_t =a(\Phi_n(\mathbf{r}),t)\Phi_n(\mathbf{r})+b(t)\bm{\epsilon}+\mathbf{r}. 
\vspace{-3pt}
\end{equation}
Substitute $\mathbf{x}_t$ with $\mathbf{x}^{(2)}_t$, Equation \ref{eq:loss_final} becomes:
\vspace{-3pt}
\begin{equation}
\small
\label{eq:loss_21}
\begin{aligned}
 &\mathcal{L}_{2,1}(\mathbf{r})=  \vert\vert F_\theta(\mathbf{x}^{(2)}_t(\bm{\epsilon}_1, \mathbf{r}),t)-f(\mathbf{x}^{(2)}_t(\bm{\epsilon}_1, \mathbf{r}),t),\bm{\epsilon}_1)\\& -F_\theta(\mathbf{x}_{t}^{(2)}(\bm{\epsilon}_2, \mathbf{r}),t)+f(\mathbf{x}_t^{(2)}(\bm{\epsilon}_2, \mathbf{r}),t),\bm{\epsilon}_2)\vert\vert_2-\lambda\vert\vert \mathbf{r}\vert\vert_2 
\end{aligned}
\end{equation}
In addition to the ending constraint for Equation \ref{eq:t_input}, we can also simplify it with the beginning constraint: Know that $\lim\limits_{t\rightarrow0}\mathbf{x}_t=\mathbf{x}_0$. Therefore, for $t\in[0,\delta]$ and $\delta\ll T$, $\mathbf{x}_t$ can be approximated with $\mathbf{x}_t^{(3)}$:
\begin{equation}
\mathbf{x}^{(3)}_t =\Phi_n(\mathbf{r}).  
\end{equation}
Substitute $\mathbf{x}_t$ with $\mathbf{x}^{(3)}_t$, Equation \ref{eq:loss_final} becomes:
\vspace{-3pt}
\begin{equation}
\small
\label{eq:loss_2_2}
\begin{aligned}
&\mathcal{L}_{2,2}(\mathbf{r}) =  \vert\vert F_\theta(\mathbf{x}^{(3)}_t(\bm{\epsilon}_1, \mathbf{r}),t)-f(\mathbf{x}^{(3)}_t(\bm{\epsilon}_1, \mathbf{r}),t),\bm{\epsilon}_1)\\& -F_\theta(\mathbf{x}_{t}^{(3)}(\bm{\epsilon}_2, \mathbf{r}),t)+f(\mathbf{x}_t^{(3)}(\bm{\epsilon}_2, \mathbf{r}),t),\bm{\epsilon}_2)\vert\vert_2-\lambda\vert\vert \mathbf{r}\vert\vert_2 
\vspace{-15pt}
\end{aligned}
\end{equation}
For simplicity, we average $\mathcal{L}_{2,1}$ and $\mathcal{L}_{2,2}$ to get our final loss for trigger refinement:
\vspace{-3pt}
\begin{equation}
\label{eq:loss}
    \mathcal{L}_{2}(\mathbf{r}) = \frac{1}{2}\mathcal{L}_{2,1}(\mathbf{r})+\frac{1}{2}\mathcal{L}_{2,2}(\mathbf{r})
\vspace{-3pt}
\end{equation}
For the overall algorithm for trigger reversion, please refer to Appendix \ref{sec:algo_reverse} for details.

\section{Backdoor Detection}

\begin{figure}[t]
    \centering
     \resizebox{0.9\linewidth}{!}{
	\includegraphics[width=0.7\linewidth]{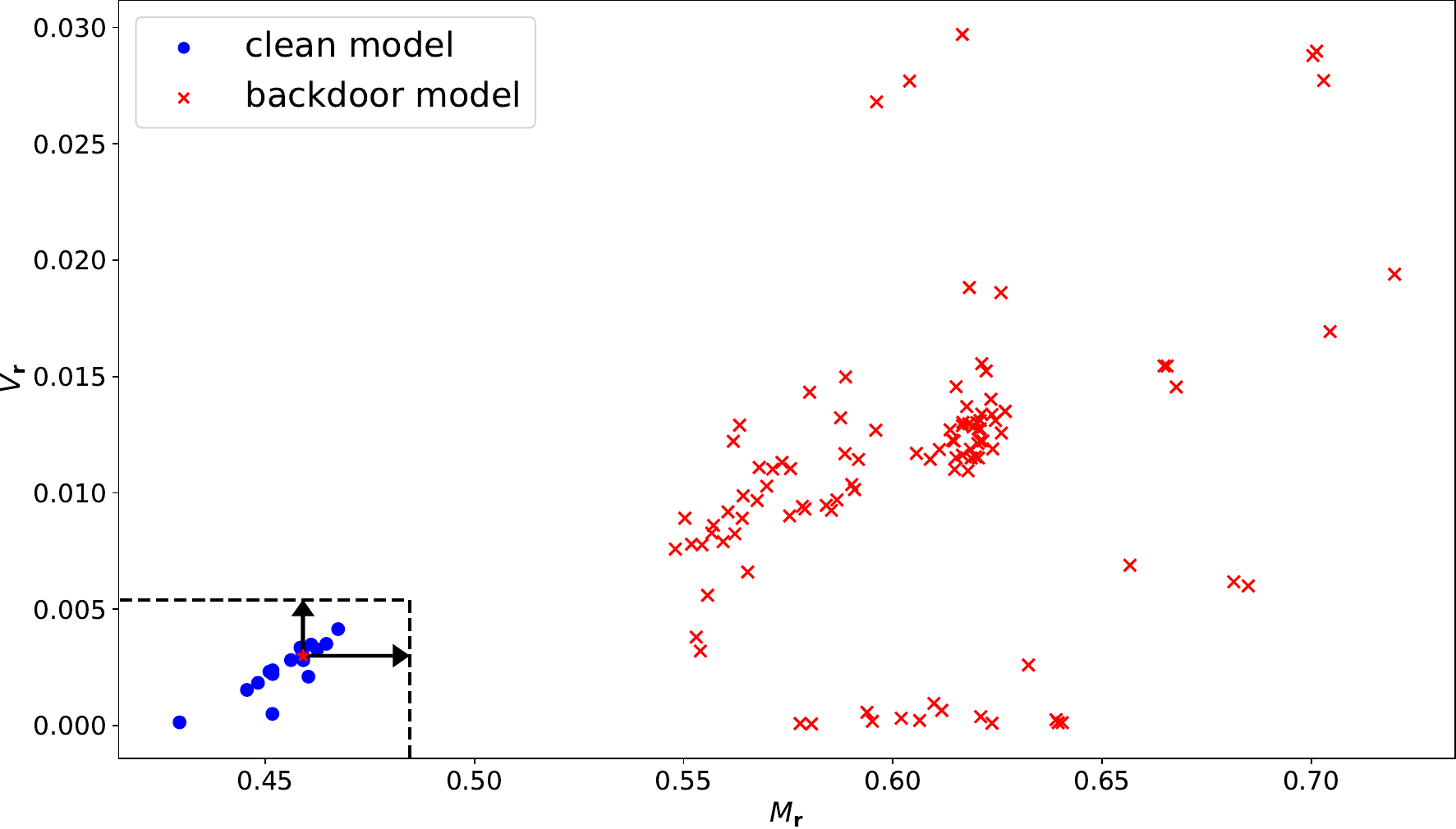}
}
 \vspace{-10pt}
    \caption{$M_\textbf{r}$ and $V_\textbf{r}$  for clean and backdoor models.}
        \vspace{-9pt}
\label{fig:model_detection}
\end{figure}

\subsection{Input Detection}

 As demonstrated in Section \ref{sec:existing_defense}, in the inference stage, because the inputs for diffusion models are sampled noises instead of natural images, current input detection methods, including \cite{chen2018detecting,zeng2021rethinking}, fail to protect diffusion models from backdoor attacks. However, if we regard the reversed trigger as the mean of the backdoor distribution, we can further detect the backdoor input from the probabilistic perspective: Note that currently, there are two distributions obtained: One is the benign distribution $\mathcal{N}(0,\mathbf{I})$, known to defenders even without defense and the other is the reversed backdoor distribution, $\mathcal{N}(\mathbf{r}, \bm{\gamma}^2)$. Here $\bm{\gamma}$ is equal to $\mathbf{I}$ for Baddiffusion and VillanDiffusion. For TrojDiff, it is co-optimized with the triggers. Given any input noise $\mathbf{\bar{\bm{\epsilon}}}$, we can calculate its probabilities in the benign or backdoor distributions, which are denoted as $\Phi_{be} (\mathbf{\bar{\bm{\epsilon}}})$ and $\Phi _{bd}(\bar{\mathbf{\bm{\epsilon}}})$, respectively. Empirically, if $\mathbf{\bar{\bm{\epsilon}}}$ is a backdoor input, $\Phi _{bd}(\mathbf{\bar{\bm{\epsilon}}})$ will be greater than $\Phi_{be} (\mathbf{\bar{\bm{\epsilon}}})$ and vice versa. Therefore, we will keep $\bm{\epsilon}$ whose $\Phi_{be} (\mathbf{\bar{\bm{\epsilon}}})\geq\Phi _{bd}(\mathbf{\bar{\bm{\epsilon}}})$ and filter out those noises with $\Phi_{be} (\mathbf{\bar{\bm{\epsilon}}})<\Phi _{bd}(\mathbf{\bar{\bm{\epsilon}}})$ because they might be backdoor inputs.

\begin{table*}[t]
    \centering
    \caption{Performance of our proposed defense against current diffusion backdoor attacks on CIFAR-10 dataset. Elijah is chosen as our baseline. The better results are in \textbf{bold}.}
    \resizebox{0.95\linewidth}{!}{
    \begin{tabular}{ccccccccccc}
    \toprule
         \multirow{2}*{Attack}&  \multirow{2}*{Defense}&\multirow{2}*{$\vert\vert \mathbf{r}-\mathbf{r}_o\vert\vert_2\downarrow$}& \multicolumn{2}{c}{Input Detection}&&\multicolumn{2}{c}{Model Detection}&&\multicolumn{2}{c}{Model Detection (BO)} \\
         \cmidrule{4-5}\cmidrule{7-8}\cmidrule{10-11}
         &   & &TPR(\%)$\uparrow$ &TNR(\%)$\uparrow$&&TPR(\%)$\uparrow$&TNR(\%)$\uparrow$&&TPR(\%)$\uparrow$&TNR(\%)$\uparrow$\\
    \midrule
    \multirow{2}*{BadDiffusion}&Elijah&32.90 & - &-&  &100.00&51.67&&68.00&21.55 \\
    & Ours&\textbf{20.69} &\textbf{100.00}&\textbf{100.00} &&\textbf{100.00}&\textbf{100.00}&&\textbf{100.00}&\textbf{100.00}\\
    \midrule
    \multirow{2}*{TrojDiff}&Elijah&22.60&-&-&&0.00&100.00&&60.00&47.50 \\
    & Ours&\textbf{4.26}&\textbf{100.00}&\textbf{100.00}&&\textbf{100.00}&\textbf{100.00}&&\textbf{100.00}&\textbf{100.00} \\
    \midrule
    \multirow{2}*{VillanDiffusion}&Elijah&43.03&-&-&&3.00&62.33&&50.00&58.33 \\
    & Ours &\textbf{30.03}&\textbf{100.00}&\textbf{100.00}&&\textbf{100.00}&\textbf{100.00}&&\textbf{100.00}&\textbf{100.00}\\
    \toprule
    \end{tabular}}
    \label{tab:main}
     \vspace{-12pt}
\end{table*}

\begin{figure*}[t]
\vspace{-1pt}
    \centering
     \resizebox{1.0\linewidth}{!}{
	\includegraphics[width=0.23\linewidth]{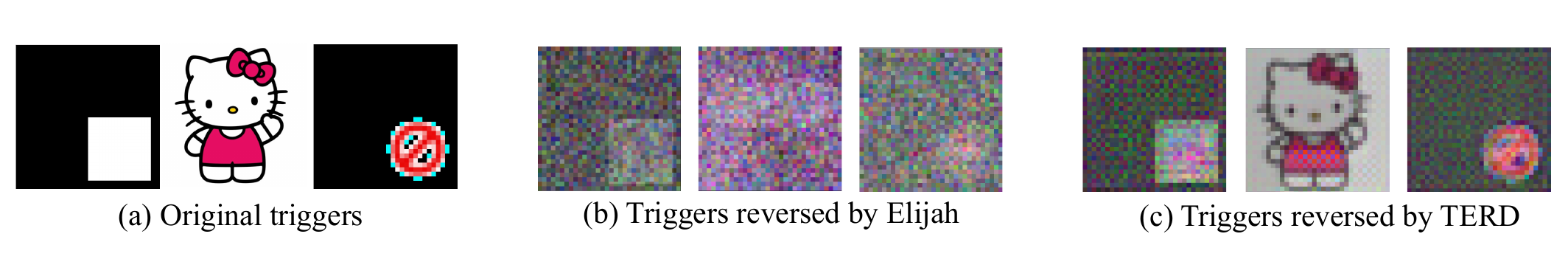}
}
 \vspace{-20pt}
\caption{Reversed results. From left to right: one of the triggers inserted by badiffusion, TrojDiff and VillanDiffusion. As can be seen, TERD more accurately reverses the triggers.}
    \vspace{-12pt}
\label{fig:reversed_trigger}
\end{figure*}
\vspace{-5pt}
\subsection{Model Detection}

In \cite{an2023remove}, they propose Elijah, the first backdoor model detection method for diffusion models. They first generate the target images with the triggers and further perform backdoor model detection with additional assumptions for target distribution. First, they assume that target images are those with high similarity. Unfortunately, this contradicts the proposition by TrojDiff, in which they demonstrate that the attacks that include multiple target images can also be applied to implant backdoors for the diffusion models. In addition, because of the discrepancy between the reversed and the original triggers, target images can not be properly generated with multiple-step generations in some hard situations. Therefore, our proposed model detection method is performed in the trigger space rather than the image space. 

Recall that in Section \ref{sec:unified}, we prove that $\mathbf{r}$ is a non-zero minimum point for the lower bound in Equation \ref{eq:elbo}. However, for the benign models, optimizing Equation \ref{eq:loss_final} will finally converge to the point that is close to a full-zero tensor because there are non-zero solutions for them. Therefore, we introduce Kullback-Leibler (KL) divergence, a metric that measures the distance between the reversed distribution $\mathcal{N}(\mathbf{r}, \bm{\gamma}^2)$ and benign distribution $\mathcal{N}(0, \mathbf{I})$. If $\mathbf{r}$ is flattened with a $n$-dimensional tensor, we can easily calculate the dimensional-wise divergence, $\mathbf{d}_{\mathbf{r}}$ between the known benign and the reversed distributions. Further, we can squeeze $\mathbf{d}_{\mathbf{r}}$ to a scalar by calculating its mean and variance over dimensions:
\vspace{-10pt}
\begin{equation}
\label{eq:mr}
\begin{aligned}
    & M_\mathbf{r}=\frac{1}{n}\sum\limits_{i=0}^{n-1}\mathbf{d}_{\mathbf{r}}[i], \\ & V_\mathbf{r}=\frac{1}{n}\sum\limits_{i=0}^{n-1}(\mathbf{d}_{\mathbf{r}}[i]-M_\mathbf{r})^2
\vspace{-15pt}
\end{aligned}
\end{equation}
For the whole-image attacks, the trigger will cause a large $M_\mathbf{r}$ because the offsets of distribution have appeared across the entire image. For the patch-based attacks, the trigger is only attached to a small region, which will lead to a large $V_\mathbf{r}$. Only the benign models can obtain low values in both $M_\mathbf{r}$ and $V_\mathbf{r}$. In Figure \ref{fig:model_detection}, we show that the backdoor and benign models can be easily detected with these extracted features. If both benign and backdoor models are available for defenders, we can train a one-layer network for model detection. We also consider a benign-only (BO) scenario, in which only benign models are accessible. We can calculate the mean and variance of $M_\mathbf{r}$ and $V_\mathbf{r}$, denoted as ($\mu_m$,$\gamma_m$) and ($\mu_v$,$\gamma_v$). According to the 3$\sigma$ criterion, any model that achieves $M_\mathbf{r}>\mu_m+3*\gamma_m$ or $V_\mathbf{r}>\mu_v+3*\gamma_v$ will be regarded as the backdoor model. 
\section{Experiment}
\begin{table*}[t]
    \centering
    \caption{Performance (\%) of our proposed defense against current diffusion backdoor attacks on high-resolution datasets.}
    \begin{tabular}{ccccccccccc}
    \toprule
           \multirow{2}*{Attack}& \multicolumn{2}{c}{Input Detection}&&\multicolumn{2}{c}{Model Detection}&&\multicolumn{2}{c}{Model Detection (BO)} \\
         \cmidrule{2-3}\cmidrule{5-6}\cmidrule{8-9}
         &TPR$\uparrow$ &TNR$\uparrow$&&TPR$\uparrow$&TNR$\uparrow$&&TPR$\uparrow$&TNR$\uparrow$\\
    \midrule
    BadDiffusion-DDPM-CelebHQ&100.00&100.00&&100.00&100.00&&100.00&100.00\\
    TrojDiff-DDPM-CelebA&100.00&100.00&&100.00&100.00&&100.00&100.00 \\
    VillanDiffusion-LDM-CelebHQ&100.00&100.00&&100.00&100.00&&100.00&100.00 \\
    \toprule
    \end{tabular}%
    \label{tab:big_dataset}
        \vspace{-5pt}
\end{table*}

\begin{table*}[t]
    \centering
    \caption{Performance (\%) of our proposed defense against current backdoor attacks on other SDE-based Models.}
    \begin{tabular}{ccccccccccc}
    \toprule
           \multirow{2}*{Model}& \multicolumn{2}{c}{Input Detection}&&\multicolumn{2}{c}{Model Detection}&&\multicolumn{2}{c}{Model Detection (BO)} \\
         \cmidrule{2-3}\cmidrule{5-6}\cmidrule{8-9}
         &TPR$\uparrow$ &TNR$\uparrow$&&TPR$\uparrow$&TNR$\uparrow$&&TPR$\uparrow$&TNR$\uparrow$\\
    \midrule
    Score-based Model&100.00&100.00&&100.00&100.00&&100.00&100.00 \\
    Consistency Model&100.00&100.00&&100.00&100.00&&100.00&100.00\\
    \toprule
    \end{tabular}%
    \label{tab:sde_based}
    \vspace{-12pt}
\end{table*}

\subsection{Experimental Settings}

\textbf{Dataset:} Our experiments are mainly performed on the CIFAR-10 \cite{krizhevsky2009learning} dataset. In Section \ref{sec:large}, we extend our experiments to large datasets, including CelebA \cite{liu2015faceattributes} and CelebA-HQ \cite{karras2017progressive}. 

\textbf{Attack:} We evaluate the performances of our defense against all known pixel-level backdoor attacks for diffusion models, including BadDiffusion, TrojDiff, and VillanDiffusion. We select the DDPM \cite{ho2020denoising} as the victim model for both BadDiffusion and TrojDiff. For VillanDiffusion, the backdoor is inserted in EDM \cite{karras2022elucidating}. To ensure a comprehensive and fair evaluation, on the CIFAR-10 dataset, we report the results that are the average of six different settings for each attack. For large datasets, all default settings from the original paper are included. Please refer to Appendix \ref{sec:config_at} for more details.

\textbf{Defense:} As far as we know, Elijah \cite{an2023remove} is the first and only existing work that specifically designs backdoor defense for diffusion models and we select it as the baseline. For its hyperparameter setting, we keep in line with the original paper. As for our proposed TERD, the iterations for trigger estimation are 3000 and 1000 for further refinement. We choose SGD as our optimizer with 0.5 learning rate which is adaptively adjusted with the cosine learning rate schedule. The trade-off coefficient $\gamma$ is set as 5e-5 for CIFAR-10 and 5e-4 for larger datasets. $\delta$ is set as $0.01T$ and the step number, $n$, for multi-step generation is set as 10. For the model detection with a neural network, we trained the model with 5 benign models and 50 backdoor models which are poisoned by the grey-box-hat setting under the BadDiffusion attack. For the benign-only (BO) backdoor detection, we calculate the threshold with 100 benign models only which are trained with the baddiffusion open-source code.

\textbf{Metrics:} To evaluate the performance of our proposed reversed engineering approach, we select the $l_2$ norm of the difference between reversed trigger $\mathbf{r}$ and the original trigger $\mathbf{r}_o$ to access the quality of the method, denoted as $\vert\vert \mathbf{r}-\mathbf{r}_o\vert\vert_2$. For the backdoor detection methods, we use TPR (True Positive Rate) and TNR (True Negative Rate): the proportion of the benign or backdoor input/model is successfully detected. For input detection, the metrics are calculated over 50000 points sampled from the benign or the backdoor distributions. For model detection, we report the results that include 100 benign models and 120 backdoor models (20 models for each of the settings). All experiments are performed on the NVIDIA A100 GPUs.

\begin{figure*}[t]
  \begin{minipage}{0.5\textwidth}
    \centering
    \captionof{table}{The effect of each component on the final performances of our proposed defense. The best results are in \textbf{bold}.}
    \resizebox{0.95\linewidth}{!}{
    \begin{tabular}{ccc|ccccccccc}
    \toprule
         \multicolumn{2}{c}{Metrics} &&TE&TR&TE+TR\\
    \midrule
    \multicolumn{2}{c}{$\vert\vert \mathbf{r}-\mathbf{r}_o\vert\vert_2\downarrow$}&&21.90&23.56 &\textbf{18.33} \\
    \midrule
    \multirow{2}*{Input Detection}&TPR(\%)&&100.00&100.00&100.00\\
    &TNR(\%)&&94.44&89.19&\textbf{100.00}\\
    \midrule
    \multirow{2}*{Model Detection}&TPR(\%)&&33.33&100.00&100.00\\
    &TNR(\%)&&88.89&77.78&\textbf{100.00}\\
    \midrule
    \multirow{2}*{Model Detection (BO)}&TPR(\%)&&33.33&100.00&100.00 \\
     &TNR(\%)&&88.89&83.33&\textbf{100.00}\\
    \toprule
    \end{tabular}}
    \label{tab:abla_overall}
  \end{minipage}
  \begin{minipage}{0.5\textwidth}
    \centering
    \includegraphics[width=0.7\textwidth]{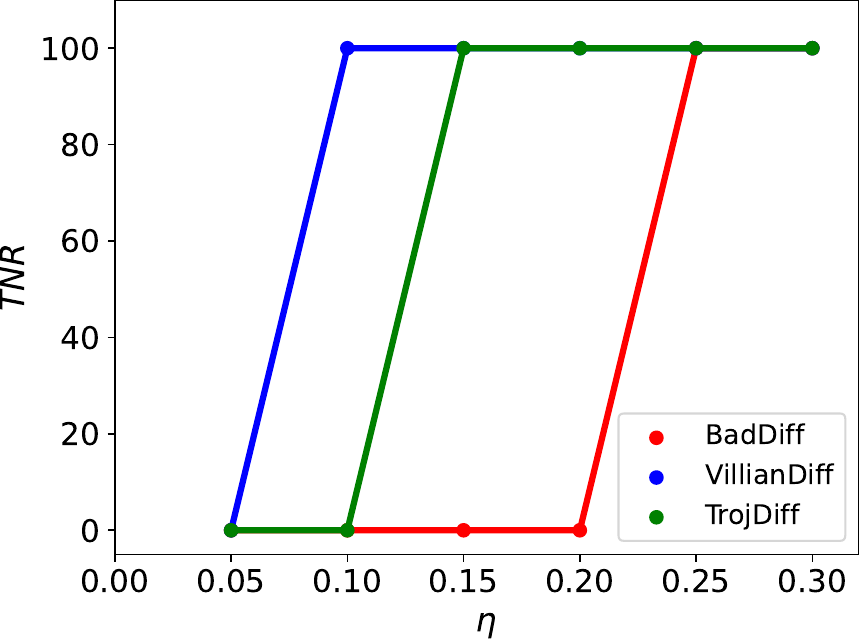}
    \vspace{-4pt}
    \caption{The performance of our proposed model detection against the adaptive attack.}
       \vspace{-15pt}
    \label{fig:aaptive}
    
  \end{minipage}
\end{figure*}
\subsection{Main Result}
\label{sec:main}
We summarize the performances of TERD against current attacks on the CIFAR-10 dataset in Table \ref{tab:main}. In addition, we compare TERD with Elijah from both the numerical results in Table \ref{tab:main} and empirical visualization in Figure \ref{fig:reversed_trigger}. First, for the reversed engineering methods, the results reveal that compared to Elijah, our proposed TERD can more accurately reverse the triggers. It is because compared to Elijah, TERD not only establishes a unified loss for trigger reversion and considers both the initial and the ending conditions of current attacks. Besides, our proposed progressively reversed strategy can help us initially estimate the trigger and improve its quality with further refinement.

Attribute to the success of our trigger reversion approach, our proposed backdoor detection method obtains 100\% TPR and TNR in all settings. From the perspective of input detection, we successfully detect the noises sampled from the backdoor distribution with the calculated probabilities. As for model detection, considering we only include one setting of the BadDiffusion attack to train the detection model, our proposed defense shows its better transferability than Elijah across different settings within the same attack and the settings across attacks. With further analysis, we find the reason is that the quality of generated images with the reversed triggers by Elijah will severely decline in some circumstances. Instead of detecting the poisoned models with the generated images, our proposed TERD performs model detection with the KL divergence of the reversed trigger. This helps TERD obtain steady performances in all settings.

\subsection{Performance on High-Resolution Dataset}
\label{sec:large}
In addition to small datasets \textit{e.g.} CIFAR-10, recent advancements in diffusion models show their outstanding performances in high-resolution image generation \cite{rombach2022high}. Unfortunately, recent studies show that backdoors can be successfully implanted even for those complex datasets \cite{chou2023backdoor}. Therefore, it is necessary to evaluate TERD on large datasets to study whether it can provide assistance for diffusion models in all situations. With the open-source code provided by current attacks, we evaluate TERD on CelebA and CelebA-HQ datasets. Since our extracted features for model detection are agnostic to the image size, we use the same detection model and the threshold adopted by the CIFAR-10 dataset. The results are summarized in Table \ref{tab:big_dataset} and for all settings, we obtain 100\% TPR and TNR. Note that the entry for attacks means the kind of attack, the victim model and the poison datasets. The results reveal that TERD is effective on high-resolution datasets and has good transferability across datasets. It means we can detect the backdoor models with TERD trained on large datasets with a detector, trained on small datasets. It can largely decrease the computation cost, considering training a diffusion model on large datasets usually requires huge computational resources.
\subsection{Transferability to SDE-based Models}
In \cite{chou2023villandiffusion}, they propose an SDE-based framework to implant a backdoor for diffusion models. Previous studies in \cite{song2020score,song2023consistency} propose that SDE can also be used to depict the dynamics of other kinds of generative models including the score-based models \cite{song2019generative} and consistency models \cite{song2023consistency}. Unfortunately, it also indicates that with some appropriate adaptations, VillanDiffusion will pose a threat not only to diffusion models but also to other models designed by similar dynamics. To study whether TERD can be applied to those models, we evaluate its performance in Table \ref{tab:sde_based}. We report the results that are the average of six configurations of VillanDiffusion and we use the same detector as this used in Section \ref{sec:main}. Surprisingly, we show that TERD can be flexibly adapted and safeguard those models. This is because TERD provides the overall framework for backdoor defense and we can instantiate its details based on different circumstances. This demonstrates the good transferability of TERD to SDE-based models and its excellent scalability even for some unknown models designed with similar principles.
\subsection{Ablation Study}
We study the effect of each component to the performance of our proposed defense. In addition, we consider defending attacks with varied trigger sizes and different poison rates. We report the results that are the averages of BadDiffusion, TrojDiff, and VillanDiffusion attacks. 

\textbf{Influence of each Component:} We compare TERD with two variants, including (1) TERD with only TE (Trigger Estimation) applied. (2) TERD with only TR (Trigger Refinement) applied on the CIFAR-10 dataset. For both of the variants, we simply substitute the loss function of the removed stage with this of the kept stage and keep other hyperparameters unchanged. As shown in Table \ref{tab:abla_overall}, although applying either TE or TR alone can yield decent performance, combining them together can obtain a more powerful defense: lower $l_2$ norm between the reversed and the original trigger, both TPR and TNR reaches 100\%. The reason is that TE estimates the target image with a surrogate distribution which might introduce randomness to the trigger reversion. And TR involves multiple forward or backward propagations through the network increasing the difficulty of optimization when it is initialized with random noises. Therefore, we propose to use TE to boost TR: by initializing noise with a rough trigger reversed by TE, the performances of TR can be further improved thus boosting the performances of both input and model detections.
 
\textbf{Trigger Size and Poison Rate}: We also investigate whether the success of TERD will be affected by the configurations of attacks. Here we consider two key factors: the size of the trigger and the poison rate. Four different settings are chosen for each factor. The minimum poison rate is set to 2\% because any value below this threshold would render the attack unsuccessful. We summarize the results in Table \ref{tab:attack_detail} of Appendix \ref{sec:config_at}. The results demonstrate that TERD obtains 100\% successful detection rates in all settings. It reveals that TERD exhibits excellent adaptability to attack with different configurations.
\subsection{Adaptive Attack}
Because we perform the backdoor detection from the distribution view, one intuitive adaptive attack is when the benign and backdoor distributions are close enough, it might bypass our proposed defense. Therefore, we introduce the hyperparameter $\eta$ ($0<\eta<1$), which scales the original trigger $\mathbf{r}_o$ to $\eta *\mathbf{r}_o$ and evaluate the performance of TERD for each settings of the attack. The TNR for model detection is summarized in Figure \ref{fig:aaptive}, which is the average of the results with a network and statistical detector. For the performances of the input detection, please refer to Figure \ref{fig:input} for details. We observe that when $\eta$ is extremely low, \textit{e.g.} $0.1$ for TrojDiff, the performance of TERD will degrade. Nevertheless, with further inspection in Table \ref{tab:clean_sample}, we find that the benign utility will be severely hurt by current attacks. This is because the backdoor and benign distributions at this time have largely overlapped. Even without TERD, the anomalies can be easily noticed by the defenders with human inspection. This illustrates the robustness of TERD to adaptive attack.

\subsection{Complexity and Time Cost}

In previous sections, we illustrate the outstanding performances of TERD in various settings. Here, we analyze the complexity of TERD to investigate whether it is practical to deploy it in real life. For our proposed reversed engineering method, the time cost is the sum of those in both stages. First, for trigger estimation, because $x_t$ can be directly represented with one equation, the computational complexity for Equation \ref{eq:loss_final} is $O(1)$. If we denote the number of iterations for the trigger estimation as $m_1$, the computational complexity for trigger estimation can be represented as $O(m_1)$. For the trigger refinement stage, we can first obtain that the complexity for obtaining $x_0$ is $O(n)$ because it needs $n$ steps of generative iterations to obtain $x_0$ and the complexity for each step is $O(1)$. Following the previous analysis for trigger estimation, we can further obtain that the overall computational complexity for the trigger refinement stage is $O(nm_2)$ ($m_2$ is the number of optimizations in the second stage.). Suming the results of both stages, the overall computational complexity for our method is $O(nm_2+m_1)$. For the analysis of the input and model detection, please refer to Appendix \ref{ap:ana_detect} for details.

In addition to the theoretical perspective, we also evaluate the time consumption with experiments. Evaluated on a single A100 GPU, we record the time consumed by TERD and the cost of training a diffusion model from scratch on the CIFAR-10 dataset in Table \ref{tab:time}. Firstly, the results indicate that compared to the training cost of diffusion models, the cost for TERD is marginal ($<1\%$). This demonstrates the cheap computational cost of TERD, which can be afforded by most defenders. Secondly, it also reveals that the detection task can be finished in less than 0.003 seconds, demonstrating our proposed method is appropriate to deploy online. It will have a negligible effect on the experience of users and can quickly finish the filtering mission even if thousands of user requests are sent to the central server.
\begin{table}[t]
    \centering
    \caption{The time cost of TERD on CIFAR-10 dataset. The time is recorded based on our experiments on a single A100 GPU.}
    \resizebox{1.0\linewidth}{!}{
    \begin{tabular}{c|ccccccccccc}
    \toprule
    Time &BadDiffusion&TrojDiff&VillanDiffusion\\
    \midrule
    Training&29h41min&45h29min&54h28min \\
    \midrule
    \midrule
    Reverse Engineering&11.13min&14.80 min&24.45min
\\
    \midrule
    Model Detection&0.0009s&0.0008s&0.0009s\\
    \midrule
     Input Detection&0.0028s&0.0025s&0.0027s\\
    \toprule
    \end{tabular}}
    \label{tab:time}
    \vspace{-20pt}
\end{table}
\section{Conclusion}
In this paper, we propose TERD, a defense framework to protect diffusion models from backdoor attacks. First, we establish a unified form for current attacks and achieve an accessible loss for reversion by applying the triangle inequality. Furthermore, we develop a two-step trigger reversion algorithm, including estimating the trigger with a substituted distribution and refining its quality with a multi-step sampler. In addition, we propose the first input detection approach by comparing probabilities across distributions and a brand new model detection method by selecting the KL divergence between the reversed and benign distributions as the metrics. We hope TERD, including the trigger reversion and backdoor detection partitions, will serve as the cornerstone to improve the backdoor robustness of diffusion models in the future.

\section*{Acknowledgements}
Yisen Wang was supported by National Key R\&D Program of China (2022ZD0160300), National Natural Science Foundation of China (62376010, 92370129), Beijing Nova Program (20230484344), and CCF-Baichuan-EB Fund.

\section*{Impact Statement}
Backdoor attacks have emerged as a significant threat to contemporary state-of-the-art diffusion models. In response, we propose the use of TERD as a defense mechanism to safeguard these models, offering the potential to enhance their overall security. Our approach is aligned with the ethical utilization of generative models, actively discouraging the generation of harmful or inappropriate content. However, it is essential to consider its environmental impact, as it may contribute to additional carbon dioxide emissions. Furthermore, it is crucial to emphasize that this paper does not intend to instill over-optimism regarding the security of diffusion models within communities. The backdoor attack, while noteworthy, is just one aspect of the potential risks faced by diffusion models. Achieving secure and trustworthy diffusion models is still a complex and ongoing journey, with many challenges ahead.

\bibliography{example_paper}
\bibliographystyle{icml2024}

\newpage
\appendix
\onecolumn
\section{The Proof of Theorem \ref{theory:1}.}
\label{ap:prof}
We first prove that the current backdoor diffusion processes are all Wiener processes in \ref{ap:prof1}. Then we further illustrate that the non-negativity of the derivative for $D_{KL}(p_t|| q_t)$.

\subsection{Wiener Processes}
\label{ap:prof1}
\textbf{TrojDiff:}
According to \cite{chen2023trojdiff}, for $\forall t \in \mathcal{Z}^+$, the relationship between $\mathbf{x}_t$ and the target image $\mathbf{x}_0$ can be formulated as:
\begin{equation}
\begin{aligned}
    \mathbf{x}_t=\sqrt{\bar{\alpha}_t}\mathbf{x}_0+\sqrt{1-\bar{\alpha}_t}\bm{\gamma\epsilon}_1+\sqrt{1-\bar{\alpha}_t}\mathbf{r} ,\quad\bm{\epsilon}_1\sim\mathcal{N}(0,\mathbf{I}).
\label{ap:prof11}
\end{aligned}
\end{equation}
where $0<\bar{\alpha}_t<1$ and it monotonically increases with $t$. $\bm{\gamma}$ is the blending coefficient and $\mathbf{r}$ denotes the trigger. For another timestep, $\forall t' \in \mathcal{Z}^+$ and $t' \leq t$, we can have the similar representation:
\begin{equation}
\begin{aligned}
    \mathbf{x}_{t'}=\sqrt{\bar{\alpha}_{t'}}\mathbf{x}_0+\sqrt{1-\bar{\alpha}_{t'}}\bm{\gamma}\bm{\epsilon}_2+\sqrt{1-\bar{\alpha}_{t'}}\mathbf{r},\quad\bm{\epsilon}_2\sim\mathcal{N}(0,\mathbf{I}).
\label{ap:prof12}
\end{aligned}
\end{equation}
It could be further re-formulized as:
\begin{equation}
\begin{aligned}
    \mathbf{x}_0= \frac{\mathbf{x}_{t'}-\sqrt{1-\bar{\alpha}_{t'}}\mathbf{r}}{\sqrt{\bar{\alpha}_{t'}}}-\frac{\sqrt{1-\bar{\alpha}_{t'}}}{\sqrt{\bar{\alpha}_{t'}}}\bm{\gamma\epsilon}_2.
\label{ap:prof13}
\end{aligned}
\end{equation}
Substitute $\mathbf{x}_0$ in Equation \ref{ap:prof11} with Equation \ref{ap:prof13}:
\begin{equation}
\begin{aligned}
     \mathbf{x}_t=\sqrt{\bar{\alpha}_t}\{{\frac{\mathbf{x}_{t'}-\sqrt{1-\bar{\alpha}_{t'}}\mathbf{r}}{\sqrt{\bar{\alpha}_{t'}}}-\frac{\sqrt{1-\bar{\alpha}_{t'}}}{\sqrt{\bar{\alpha}_{t'}}}\bm{\gamma}\bm{\epsilon}_2}\}+\sqrt{1-\bar{\alpha}_t}\bm{\gamma}\bm{\epsilon}_1+\sqrt{1-\bar{\alpha}_t}\mathbf{r}.
\label{ap:prof14}
\end{aligned}
\end{equation}
Because $\bm{\epsilon}_1$ is independent of $\bm{\epsilon}_2$, we can combine them together and introduce a new variable $\bm{\epsilon}$:
\begin{equation}
\begin{aligned}
    \mathbf{x}_t=\sqrt{\frac{\bar{\alpha}_{t}}{\bar{\alpha}_{t'}}}\mathbf{x}_{t'}-\sqrt{\frac{\bar{\alpha}_{t}}{\bar{\alpha}_{t'}}}\sqrt{1-\bar{\alpha}_{t'}}\mathbf{r}+\sqrt{1-\bar{\alpha}_{t}}\mathbf{r}+\sqrt{1-\frac{\bar{\alpha}_{t}}{\bar{\alpha}_{t'}}}\bm{\gamma}\bm{\epsilon},\quad \bm{\epsilon}\sim\mathcal{N}(0,\mathbf{I}).
\label{ap:prof15}
\end{aligned}
\end{equation}          
A more symmetric form is
\begin{equation}
\begin{aligned}
    \frac{\mathbf{x}_t}{\bm{\gamma}\sqrt{\bar{\alpha}_{t}}}-\frac{\sqrt{1-\bar{\alpha}_{t}}}{\bm{\gamma}\sqrt{\bar{\alpha}_t}}\mathbf{r}=\frac{\mathbf{x}_{t'}}{\bm{\gamma}\sqrt{\bar{\alpha}_{t'}}}-\frac{\sqrt{1-\bar{\alpha}_{t'}}}{\bm{\gamma}\sqrt{\bar{\alpha}_{t'}}}\mathbf{r}+\sqrt{\frac{1}{\bar{\alpha}_t}-\frac{1}{\bar{\alpha}_{t'}}}\bm{\epsilon}.
\label{ap:prof16}
\end{aligned}
\end{equation}         
We can replace it with new variables:
\begin{equation}
\begin{cases}
    s_t &= \frac{1}{\bar{\alpha}_t}-\frac{1}{\bar{\alpha}_{0}}, \quad t\in\mathbb{Z}^+\\
    \mathbf{y}_{s_t} &= \frac{\mathbf{x}_t}{\bm{\gamma}\sqrt{\bar{\alpha}_{t}}}-\frac{\sqrt{1-\bar{\alpha}_{t}}}{\bm{\gamma}\sqrt{\bar{\alpha}_t}} \mathbf{r} -\{{\frac{\mathbf{x}_0}{\bm{\gamma}\sqrt{\bar{\alpha}_{0}}}-\frac{\sqrt{1-\bar{\alpha}_{0}}}{\bm{\gamma}\sqrt{\bar{\alpha}_0}} \mathbf{r}}\}.\\
\end{cases}
\end{equation}
For all: $s_T>s_{T-1}>\cdot\cdot\cdot>s_{0}=0$, and
\begin{equation}
\begin{cases}
\label{ap:prof17}
    \mathbf{y_{0}} = 0\\
    \mathbf{y_{s'}}-\mathbf{y_{s}} = \sqrt{s'-s}\bm{\epsilon}, \quad\bm{\epsilon}\sim\mathcal{N}(0,\mathbf{I}),\quad s'>s.\\
\end{cases}
\end{equation}  
It proves that $\mathbf{y_{s'}}$ is a Wiener process.

\textbf{BadDiffusion}: 
According to \cite{chou2023backdoor}, for $\forall t \in \mathcal{Z}^+$,  the relationship between $\mathbf{x}_t$ and the target image $\mathbf{x}_0$ can be defined as:
\begin{equation}
\begin{aligned}
\label{ap:prof18}
    \mathbf{x}_t=\sqrt{\bar{\alpha}}_t \mathbf{x}_0+\sqrt{1-\bar{\alpha}_t}\bm{\epsilon}_1+(1-\sqrt{\bar{\beta}_t})\mathbf{r},\quad\bm{\epsilon}_1\sim\mathcal{N}(0,\mathbf{I}).
\end{aligned}
\end{equation}
Here we share the same symbolic meanings with TrojDiff. In \cite{chou2023backdoor}, they set: $\bar{\beta}_t = \bar{\alpha}_t$. However, here we consider a more general case: $\{\bar{\beta}_t\}_{t=1}^T$ could be a different sequence with $\{\bar{\alpha}_t\}_{t=1}^T$. For $t'\in \mathcal{Z}^+$ and $t' \leq t$, this formulation becomes:
\begin{equation}
\begin{aligned}
\label{ap:prof19}
    \mathbf{x}_{t'}=\sqrt{\bar{\alpha}_{t'}}\mathbf{x}_0+\sqrt{1-\bar{\alpha}_{t'}}\bm{\epsilon}_2+(1-\sqrt{\bar{\beta}_{t'}})\bm{r},\quad\bm{\epsilon}_2\sim\mathcal{N}(0,\mathbf{I}).
\end{aligned}
\end{equation}
Similar to TrojDiff, we can re-formulized this equation:
\begin{equation}
\begin{aligned}
\label{ap:prof110}
    \mathbf{x}_0 = \frac{\mathbf{x}_{t'}-(1-\sqrt{\bar{\beta}_{t'}})\mathbf{r}}{\sqrt{\bar{\alpha}_{t'}}}-\frac{\sqrt{1-\bar{\alpha}_{t'}}}{\sqrt{\bar{\alpha}_{t'}}}\bm{\epsilon}_2
\end{aligned}
\end{equation}
Substitute $\mathbf{x}_0$ in Equation \ref{ap:prof18} with Equation \ref{ap:prof110}, we get
\begin{equation}
\begin{aligned}
\label{ap:prof111}
    \mathbf{x}_t=\sqrt{\bar{\alpha}_t}\{\frac{\mathbf{x}_{t'}-(1-\sqrt{\bar{\beta}_{t'}})\mathbf{r}}{\sqrt{\bar{\alpha}_{t'}}}-\frac{\sqrt{1-\bar{\alpha}_{t'}}}{\sqrt{\bar{\alpha}_{t'}}}\bm{\epsilon}_2\}+\sqrt{1-\bar{\alpha}_t}\bm{\epsilon}_1+(1-\sqrt{\bar{\beta}_t})\mathbf{r}
\end{aligned}
\end{equation}
Simplify Equation \ref{ap:prof111} and combine $\bm{\epsilon}_1$ and $\bm{\epsilon}_2$ together with $\bm{\epsilon}$, we get:
\begin{equation}
\begin{aligned}
\label{ap:prof112}
    \mathbf{x}_t=\sqrt{\frac{\bar{\alpha}_{t}}{\bar{\alpha}_{t'}}}\mathbf{x}_{t'}-\sqrt{\frac{\bar{\alpha}_{t}}{\bar{\alpha}_{t'}}}(1-\sqrt{\bar{\beta}_{t'}})\mathbf{r}+(1-\sqrt{\bar{\beta}_t})\mathbf{r}+\sqrt{1-\frac{\bar{\alpha}_{t}}{\bar{\alpha}_{t'}}}\bm{\epsilon}\quad \bm{\epsilon}\sim \mathcal{N}(0,\mathbf{I})
\end{aligned}
\end{equation}
A more symmetric form is
\begin{equation}
\begin{aligned}
\label{ap:prof113}
    \frac{\mathbf{x}_t}{\sqrt{\bar{\alpha}_{t}}}-\frac{1-\sqrt{\bar{\beta}_{t}}}{\sqrt{\bar{\alpha}_t}}\mathbf{r}=\frac{\mathbf{x}_{t'}}{\sqrt{\bar{\alpha}_{t'}}}-\frac{1-\sqrt{\bar{\beta}_{t'}}}{\sqrt{\bar{\alpha}_{t'}}}\mathbf{r}+\sqrt{\frac{1}{\bar{\alpha}_t}-\frac{1}{\bar{\alpha}_{t'}}}\bm{\epsilon}
\end{aligned}
\end{equation}
Replace it with new variables, $s_t$ and $\mathbf{y}_{s_t}$:
\begin{equation}
\begin{cases}
    s_t &= \frac{1}{\bar{\alpha}_t}-\frac{1}{\bar{\alpha}_{0}}, \quad t\in\mathbb{Z}^+\\
    \mathbf{y}_{s_t} &= \frac{\mathbf{x}_t}{\sqrt{\bar{\alpha}_{t}}}-\frac{1-\sqrt{\bar{\beta}_{t}}}{\sqrt{\bar{\alpha}_t}}\mathbf{r} -\{\frac{\mathbf{x}_0}{\sqrt{\bar{\alpha}_{0}}}+\frac{1-\sqrt{\bar{\beta}_{0}}}{\sqrt{\bar{\alpha}_0}}\mathbf{r}\}\\
\end{cases}
\end{equation}
Thus for all: $s_T>s_{T-1}>\cdot\cdot\cdot>s_{0}=0$, we have
\begin{equation}
\begin{cases}
\label{ap:prof114}
    \mathbf{y}_{0} = 0\\
    \mathbf{y}_{s'}-\mathbf{y}_{s} = \sqrt{s'-s}\bm{\epsilon},\quad\bm{\epsilon}\sim\mathcal{N}(0,\mathbf{I}),\quad s'>s.\\
\end{cases}
\end{equation} 
Thus we prove that $\mathbf{y}_{s'}$ is a Wiener process.

\textbf{VillianDiffusion}: It serves as a continuous version of BadDiffusion which means $t$ is extended to $[0,+\infty)$ and $\alpha_t$, $\beta_t$ are assumed to be a continuous functions, where $\lim\limits_{t \to \infty }\bar{\alpha}_t =0$, $\lim\limits_{t \to \infty }\bar{\beta}_t =0$ and $\lim\limits_{t \to 0
}\bar{\beta}_t =1$. Therefore, we could extend $s$ to $[0,+\infty)$ as well and prove VillianDiffusion is also a Wiener process similarly.

\subsection{Proof \ref{theory:1}}
\begin{equation}
\begin{aligned}
    \frac{\partial D_{KL}(\mathbf{p}_s||\mathbf{q}_s)}{\partial s} &= \frac{\partial}{\partial s}\int \mathbf{p}(\mathbf{y}_s)log\frac{\mathbf{p}(\mathbf{y}_s)}{\mathbf{q}(\mathbf{y}_s)}d\mathbf{y} \\
    &=\int \frac{\partial\mathbf{p}(\mathbf{y}_s)}{\partial s}log \frac{\mathbf{p}(\mathbf{y}_s)}{\mathbf{q}(\mathbf{y}_s)} d\mathbf{y}+ \int\frac{\partial\mathbf{p}(\mathbf{y}_s)}{\partial s}d\mathbf{y}
    +\int \frac{\partial\mathbf{q}(\mathbf{y}_s)}{\partial s} \frac{\mathbf{p}(\mathbf{y}_s)}{\mathbf{q}(\mathbf{y}_s)}d\mathbf{y}
\end{aligned}
\end{equation}
Then if we assume $\mathbf{p}(\mathbf{y}_s)$ and $\mathbf{q}(\mathbf{y}_s)$ are smooth and fast decaying, then $\int \frac{\partial\mathbf{p}(\mathbf{y}_s)}{\partial s}=0$ and $\frac{\partial\mathbf{q}(\mathbf{y}_s)}{\partial s}=0$. In other words, the integration of second term is $0$. Then
\begin{equation}
\begin{aligned}
\label{ap:prof21}
    \frac{\partial D_{KL}(\mathbf{p}_s||\mathbf{q}_s)}{\partial s} &=\int \frac{\partial\mathbf{p}(\mathbf{y}_s)}{\partial s}log \frac{\mathbf{p}(\mathbf{y}_s)}{\mathbf{q}(\mathbf{y}_s)}d\mathbf{y}
    +\int \frac{\partial\mathbf{q}(\mathbf{y}_s)}{\partial s} \frac{\mathbf{p}(\mathbf{y}_s)}{\mathbf{q}(\mathbf{y}_s)}d\mathbf{y}
\end{aligned}
\end{equation}
We know for Wiener process $\mathbf{y}_s$, both $\mathbf{q}(\mathbf{y}_s)$ and $\mathbf{p}(\mathbf{y}_s)$ satisfy:
\begin{equation}
\begin{aligned}
\label{ap:prof22}
    \frac{\partial \mathbf{p}(\mathbf{y}_s)}{\partial s}=\frac{1}{2} \frac{\partial^2\mathbf{p}(\mathbf{y}_s)}{\partial \mathbf{y}^2_s}     \quad \quad
    \frac{\partial \mathbf{q}(\mathbf{y}_s)}{\partial s}=\frac{1}{2} \frac{\partial^2\mathbf{q}(\mathbf{y}_s)}{\partial \mathbf{y}^2_s}
\end{aligned}
\end{equation}
Substitute $\frac{\partial \mathbf{p}(\mathbf{y}_s)}{\partial s}$ and $\frac{\partial \mathbf{q}(\mathbf{y}_s)}{\partial s}$ in Equation \ref{ap:prof21} with Equation \ref{ap:prof22}, we get
\begin{equation}
\begin{aligned}
    \frac{\partial D_{KL}(\mathbf{p}_s||\mathbf{q}_s)}{\partial s} =
    \frac{1}{2} \int \frac{\partial^2\mathbf{p}(\mathbf{y}_s)}{\partial \mathbf{y}^2_s}log \frac{\mathbf{p}(\mathbf{y}_s)}{\mathbf{q}(\mathbf{y}_s)}d\mathbf{y}+\int \frac{\partial^2\mathbf{q}(\mathbf{y}_s)}{\partial \mathbf{y}^2_s} \frac{\mathbf{p}(\mathbf{y}_s)}{\mathbf{q}(\mathbf{y}_s)}d\mathbf{y}
\end{aligned}
\end{equation}
Using integration by parts and 
\begin{equation}
    \frac{\partial \mathbf{p}(\mathbf{y}_s)}{\partial \mathbf{y}_s} = \frac{1}{\mathbf{p}(\mathbf{y}_s)}\frac{\partial log \mathbf{p}(\mathbf{y}_s)}{\partial \mathbf{y}_s} 
\end{equation}, it becomes
\begin{align}
    \frac{\partial D_{KL}(\mathbf{p}_s||\mathbf{q}_s)}{\partial s} &= -\frac{1}{2}\int \bigg(\frac{\partial \mathbf{p}(\mathbf{y}_s)}{\partial \mathbf{y}_s}\frac{\partial log \frac{\mathbf{p}(\mathbf{y}_s)}{\mathbf{q}(\mathbf{y}_s)}}{\partial \mathbf{y}_s}+ \frac{\partial \mathbf{q}(\mathbf{y}_s)}{\partial \mathbf{y}_s} \frac{\partial \frac{\mathbf{p}(\mathbf{y}_s)}{\mathbf{q}(\mathbf{y}_s)}}{\partial \mathbf{y}_s}d\mathbf{y}\bigg)d\mathbf{y} \\
    &= -\frac{1}{2} \int \bigg(\mathbf{p}(\mathbf{y}_s)\frac{\partial log \mathbf{p}(\mathbf{y}_s)}{\partial \mathbf{y}_s}\frac{\partial log\frac{\mathbf{p}(\mathbf{y}_s)}{\mathbf{q}(\mathbf{y}_s)} }{\partial \mathbf{y}_s} + \mathbf{q}(\mathbf{y}_s)\frac{\partial log \mathbf{q}(\mathbf{y}_s)}{\partial \mathbf{y}_s}\frac{\mathbf{p}(\mathbf{y}_s)}{\mathbf{q}(\mathbf{y}_s)}\frac{\partial log \frac{\mathbf{p}(\mathbf{y}_s)}{\mathbf{q}(\mathbf{y}_s)}}{\partial \mathbf{y}_s}\bigg)d\mathbf{y}  \\
    &= -\frac{1}{2} \int \mathbf{p}(\mathbf{y}_s) \frac{\partial log\frac{\mathbf{p}(\mathbf{y}_s)}{\mathbf{q}(\mathbf{y}_s)} }{\partial \mathbf{y}_s}\bigg(  \frac{1}{\mathbf{p}(\mathbf{y}_s)}\frac{\partial \mathbf{p}(\mathbf{y}_s)}{\partial \mathbf{y}_s}+ \frac{1}{\mathbf{q}(\mathbf{y}_s)}\frac{\partial \mathbf{q}(\mathbf{y}_s)}{\partial \mathbf{y}_s}\bigg)d\mathbf{y}  \\
    &=-\frac{1}{2} \int \mathbf{p}(\mathbf{y}_s) \bigg(\frac{\partial log \frac{\mathbf{p}(\mathbf{y}_s)}{\mathbf{q}(\mathbf{y}_s)}}{\partial \mathbf{y}_s}\bigg)^2 d\mathbf{y} \\
    &= -\frac{1}{2} \mathbb{E}\bigg[\bigg(\frac{\partial log \frac{\mathbf{p}(\mathbf{y}_s)}{\mathbf{q}(\mathbf{y}_s)}}{\partial \mathbf{y}_s}\bigg)^2\bigg]
\end{align}
It is Fisher information. And we know 
\begin{equation}
\begin{aligned}
    D_{F}(p_{t}||q_{t}) = \mathbb{E}\bigg[\bigg(\frac{\partial log \frac{\mathbf{p}(\mathbf{y}_s)}{\mathbf{q}(\mathbf{y}_s)}}{\partial \mathbf{y}_s}\bigg)^2\bigg] \geq 0 
\end{aligned}
\end{equation}
Therefore,
\begin{equation}
\begin{aligned}
     \frac{\partial D_{KL}(\mathbf{p}_s||\mathbf{q}_s)}{\partial s} &= -\frac{1}{2}D_{F}(\mathbf{p}_s||\mathbf{q}_s) \leq 0
\end{aligned}
\end{equation}

\section{Algorithm for TERD}

\subsection{Trigger Reversion}
\label{sec:algo_reverse}

\begin{algorithm}[H]
\caption{Trigger reversion.} 
\label{alg:algorithm1}
\begin{algorithmic}[1]
\STATE \textbf{Input:} Diffusion model $F_\theta$, random initialize $\mathbf{r}$, iteration $e_1$ $e_2$, learning rate $\eta$, $n$-step sampler $\Phi_n(\cdot)$, trade-off coefficient $\lambda$, the substituted distribution $\hat{p}_{prior}$. 
\STATE \textbf{Output:} Reversed trigger $\mathbf{r}$.
\FOR{$i\leftarrow 1,\ldots,e_1$} 
\STATE Init $\hat{\mathbf{x}}_0$ from $\hat{p}_{prior}$
\STATE Sample $t$ from $\mathcal{U}[T-\delta,T]$
\STATE Sample $\bm{\epsilon}_1 $ ,$\bm{\epsilon}_2$ from $\mathcal{N}(\mathbf{0},\mathbf{I})$
\STATE Derive $\mathbf{x}_t^1(\bm{\epsilon}_1, \mathbf{r}),\mathbf{x}_t^1(\bm{\epsilon}_2$, $\mathbf{r})$ with $\hat{\mathbf{x}}_0$
\STATE $\mathbf{r} \gets \mathbf{r} - \eta \nabla_\mathbf{r} \mathcal{L}_1(\mathbf{r})$
\COMMENT{Equation \ref{eq:loss_pi}}
\ENDFOR
\FOR {$j\leftarrow 1,\ldots,e_2$} 
\STATE $\mathbf{x}_0 \leftarrow \Phi_n(\mathbf{r})$
\STATE Sample $t_1$ from $\mathcal{U}[T-\delta,T]$
\STATE Sample $\bm{\epsilon}_1 $, $\bm{\epsilon}_2$ from $\mathcal{N}(\mathbf{0},\mathbf{I})$
\STATE Derive $\mathbf{x}_t^2(\bm{\epsilon}_1, \mathbf{r})$, $\mathbf{x}_t^2(\bm{\epsilon}_2,\mathbf{r)}$ with $\mathbf{x}_0$
\STATE Sample $t_2$ from $\mathcal{U}[0,\delta]$
\STATE Sample $\bm{\epsilon}_1$, $\bm{\epsilon}_2$ from $\mathcal{N}(\mathbf{0},\mathbf{I})$
\STATE Derive $\mathbf{x}_t^3(\bm{\epsilon}_1, \mathbf{r})$, $\mathbf{x}_t^3(\bm{\epsilon}_2,\mathbf{r)}$ with $\mathbf{x}_0$
\STATE $\mathbf{r} \gets \mathbf{r} - \eta \nabla_\mathbf{r} \mathcal{L}_2(\mathbf{r})$
\COMMENT{Equation \ref{eq:loss}}
\ENDFOR
\end{algorithmic}
\end{algorithm}

\subsection{Input detection}

\begin{algorithm}[H]
\caption{Input detection.} 
\label{alg:algorithm2}
\begin{algorithmic}[1]
\STATE \textbf{Input:} Input noise $\bm{\bar{\epsilon}}$, potential backdoor distribution $\mathcal{N}(\mathbf{r}, \bm{\gamma}^2)$.
\STATE \textbf{Output:} $\Phi_{bd} (\mathbf{\bar{x}}) \leq \Phi_{be} (\mathbf{\bar{x}})$.
\COMMENT{1 means $\bm{\bar{\epsilon}}$ is a clean input, otherwise is a backdoor input.}
\STATE $\Phi_{be} (\bm{\bar{\epsilon}}) \leftarrow$ The probability of $\bm{\bar{\epsilon}}$ in distribution $\mathcal{N}(0,\mathbf{I})$
\STATE $\Phi_{bd} (\bm{\bar{\epsilon}}) \leftarrow$ The probability of $\bm{\bar{\epsilon}}$ in distribution $\mathcal{N}(\mathbf{r}, \bm{\gamma}^2)$
\end{algorithmic}
\end{algorithm}

\subsection{Model Detection}

\begin{algorithm}[H]
\caption{Feature extraction of model detection.} 
\label{alg:algorithm3}
\begin{algorithmic}[1]
\STATE \textbf{Input:} Input model $\theta$.
\STATE \textbf{Output:} $M_\mathbf{r}$, $V_\mathbf{r}$.
\STATE $\mathbf{r} \leftarrow  Trigger Reversion(\theta)$
\STATE $\mathbf{d}_{\mathbf{r}} \leftarrow$ KL divergence between $\mathcal{N}(\mathbf{r}, \bm{\gamma}^{2})$ and $\mathcal{N}(0,\mathbf{I})$
\STATE $M_\mathbf{r} \leftarrow \frac{1}{n}\sum\limits_{i=0}^{n-1}\mathbf{d}_{\mathbf{r}}[i]$ 
\STATE $V_\mathbf{r} \leftarrow \frac{1}{n}\sum\limits_{i=0}^{n-1}(\mathbf{d}_{\mathbf{r}}[i]-M_\mathbf{r})^2$
\end{algorithmic}
\end{algorithm}

\begin{algorithm}[H]
\caption{Model detection via a network} 
\label{alg:algorithm4}
\begin{algorithmic}[1]
\STATE \textbf{Input:} $K$ models for training: $\{M_i\}_{i=1}^K$, 
and its label: $\{y_i\}_{i=1}^K$, epoch e, learning rate $\eta$, unknown model $\phi$.
\STATE \textbf{Output:} $C_\theta(\mathbf{f}_\phi)$.
\STATE  $\mathcal{D}_{train}\leftarrow \{ExtractFeature(M_i),y_i\}_{i=1}^K$ 
\STATE Randomly init classifier $C_\theta$
\FOR{$j\leftarrow 1, \ldots, e$}
\STATE $\theta \leftarrow \theta - \eta \cdot \nabla_\theta \mathcal{L}(\theta, \mathcal{D}_{train})$
\ENDFOR
\STATE $\mathbf{f}_\phi \leftarrow ExtractFeature(\phi)$
\end{algorithmic}
\end{algorithm}

\begin{algorithm}[H]
\caption{Model detection with only benign models} 
\label{alg:algorithm5}
\begin{algorithmic}[1]
\STATE \textbf{Input:} $K$ benign models: $\{B_i\}_{i=1}^K$, unknown model $\phi$.
\STATE \textbf{Output:} $\phi$ is a clean model or not.
\STATE $\mathcal{M}\leftarrow\{ExtractFeature(B_i)[M_{\mathbf{r}}]\}_{i=1}^K$
\STATE $\mathcal{V}\leftarrow\{ExtractFeature(B_i)[V_{\mathbf{r}}]\}_{i=1}^K$
\STATE $\psi_m \leftarrow$ $mean(\mathcal{M})+ 3* std(\mathcal{M})$
\STATE $\psi_v \leftarrow$ $mean(\mathcal{V})+ 3* std(\mathcal{V})$
\STATE $m$, $v \leftarrow  Extract Feature(\phi$)
\IF{$m > \psi_m$ or $v > \psi_v$}
\STATE $\phi$ is a backdoor model
\ELSE 
\STATE $\phi$ is a clean model
\ENDIF
\end{algorithmic}
\end{algorithm}

\newpage

\section{Detailed Configurations for Backdoor Attacks}
\label{sec:config_at}

\begin{table*}[h]
    \centering
    \caption{Patterns of triggers and target images for Baddifusion and VillianDiffusion attacks.}
    \begin{tabular}{c c|c c c|c |c}
    \toprule
        \multicolumn{5}{c|}{CIFAR-10 (32 × 32)}&\multicolumn{2}{c}{CelebA-HQ (256 × 256)}\\
        \midrule
        \multicolumn{2}{c|}{Triggers}&\multicolumn{3}{|c|}{Targets}& Triggers&Targets\\
        \midrule
        Grey Box& Stop Sign&Corner&Shoe&Hat&Eyeglasses&Cat\\
        \includegraphics[width=2cm,height=2cm]{Figure/box.pdf}
&\includegraphics[width=2cm,height=2cm]{Figure/stop_sign.pdf}&\includegraphics[width=2cm,height=2cm]{Figure/corn.pdf}&\includegraphics[width=2cm,height=2cm]{Figure/shoe.pdf}&\includegraphics[width=2cm,height=2cm]{Figure/hat.pdf}&\includegraphics[width=2cm,height=2cm]{Figure/glass.pdf}&\includegraphics[width=2cm,height=2cm]{Figure/cat.pdf}\\    
    \toprule
    \end{tabular}
    \label{tab:ex_setting1}
\end{table*}

\begin{table*}[h]
    \centering
    \caption{Patterns of triggers and target images for TrojDiff attack on both CIFAR-10 and CelebA datasets.}
    \begin{tabular}{c c|c c c}
    \toprule
        \multicolumn{2}{c|}{Triggers}&\multicolumn{3}{|c}{Targets}\\
        \midrule
        Patch-based &Blend-based &In-D2D attack&Out-D2D attack&D2l attack\\
        \includegraphics[width=2cm,height=2cm]{Figure/white3.pdf}
        &\includegraphics[width=2cm,height=2cm]{Figure/hello_kitty.pdf}
&\includegraphics[width=2cm,height=2cm]{Figure/d2din.pdf}&\includegraphics[width=2cm,height=2cm]{Figure/d2dout.pdf}&\includegraphics[width=2cm,height=2cm]{Figure/mickey.pdf}\\    
    \toprule
    \end{tabular}
    \label{tab:ex_setting2}
\end{table*}

\textbf{Baddifusion}:
Following the settings of backdoor attacks in \cite{chou2023backdoor}, two triggers and three target images (6 combinations) are considered for the CIFAR-10 dataset. For the experiments on the CelebA-HQ dataset, we adopt the default setting: eyeglasses trigger and cat target to implant the backdoor. The detailed patterns of them are illustrated in Table \ref{tab:ex_setting1}.
To save the computational cost, the backdoor is implanted by fine-tuning with the Adam optimizer. For the CIFAR-10 dataset, the learning rate is 2e-4 and the batch size is 128. For the CelebA-HQ dataset, the learning rate and batch size are 8e-5 and 64, respectively.

\textbf{TrojDiff}:
We include all settings of TrojDiff and blend all triggers and the sampled Gaussian noises with a coefficient of 0.6. As shown in Table \ref{tab:ex_setting2}, the image of HelloKitty is chosen as the trigger for the blend-based attack, while the classific checkerboard trigger is selected for the patch-based attack. As for target selections, we include all attack scenarios in TrojDiff with target images from in-domain images (In-D2D), out-domain images (Out-D2D), and an individual image (D2I). In the In-D2D setting, the target class corresponds to class 7, which translates to ``horse'' on the CIFAR-10 dataset and ``faces with heavy makeup, mouth slightly open, smiling'' in CelebA. For the Out-D2D setting, the handwritten number ``7'' extracted from the MNIST dataset serves as the target. As for the Out-D2I setting, the target image is a single image, the Mickey Mouse image. Details are shown in Table \ref{tab:ex_setting2} and the images are resized to different sizes according to the resolutions of the datasets. For the hyperparameter configurations, we employ the Adam optimizer with a learning rate of 0.0002 to fine-tune the pre-trained diffusion models to implant backdoor attacks. The decay rate of the Exponential Moving Average (EMA) is set as 0.9999 and the batch size is set to 128 which follows the original paper. 

\textbf{VillianDiffusion}:
As for VillianDiffusion, triggers and target images for CIFAR10 and CelebA-HQ are the same as those of Baddifusion, shown in Table \ref{tab:ex_setting1}. We fine-tune the pre-trained EDM on the CIFAR-10 dataset with learning rate 2e-4 and 128 batch size for 200000 iterations. For the experiments on the CELEBA-HQ dataset, we insert backdoors for LDM with the open-source code provided by VillianDiffusion.

\newpage

\section{Performances of TERD with varied attack configurations}

\label{sec:config_at}

\begin{table}[H]
\caption{The performances of our proposed defense with attacks of different trigger sizes and poison rates. 
}
\vspace{-5pt}
\label{tab:attack_detail}
\vskip 0.10in
\centering
\subtable[trigger size]{
\label{trigger_size}
\resizebox{0.45\linewidth}{!}{
\begin{tabular}{ccccccccccc}
    \toprule
           Trigger & Input&Model &Model \\
            Size &Detection& Detection& Detection (BO) \\
    \midrule
    $4\times4$&100.00&100.00&100.00\\
    $8\times8$&100.00&100.00 &100.00\\
    $11\times11$&100.00&100.00&100.00 \\
    $14\times14$&100.00&100.00&100.00 \\
    \toprule
    \end{tabular}}
}
\subtable[poison rate]{
\resizebox{0.45\linewidth}{!}{
    \begin{tabular}{ccccccccccc}
    \toprule
           Poison  & Input&Model &Model \\
            Rate &Detection& Detection& Detection (BO) \\
    \midrule
    $2\%$&100.00&100.00&100.00\\
    $5\%$&100.00&100.00&100.00\\
    $10\%$&100.00&100.00&100.00\\
    $20\%$& 100.00&100.00&100.00\\
    
    \toprule
    \end{tabular}}
}
\vspace{-15pt}
\end{table}

\section{Computational Analysis for the Detection Method}
\label{ap:ana_detect}
\textbf{Model detection:} The computational cost for our proposed model detection method is mainly caused by computing the metrics $M_r$ and $N_r$.
We can first assume that the reversed trigger $\mathbf{r}\in\mathbb{R}^{3\times k\times k}$.  According to the formulation of KL divergence, the computational overhead for computing $d_r$ is only proportional to the dimension of $\mathbf{r}$, and can be formulated as $O(3k^2)$. Furthermore, according to the formulation in Equation \ref{eq:mr}, the computational complexity for calculating $M_r$ and $V_r$ is also $O(3k^2)$. Therefore, the overall computational complexity is $O(3k^2)$.

\textbf{Input detection:} Similar to the analysis for model detection, we can also assume that the reversed trigger $\mathbf{r}\in\mathbb{R}^{3\times k\times k}$ . According to the probability density function of the multivariate Gaussian and the independence across dimensions. The computational complexity for calculating the probability in the backdoor or benign distribution are both $O(3k^2)$. Therefore, summing them together is also $O(3k^2)$.

\section{The Performances of TERD against the adaptive attack }
\label{sec:terd_ada}

\begin{figure}[H]
    \centering
     \subfigure[TPR]	{\includegraphics[width=0.4\linewidth]{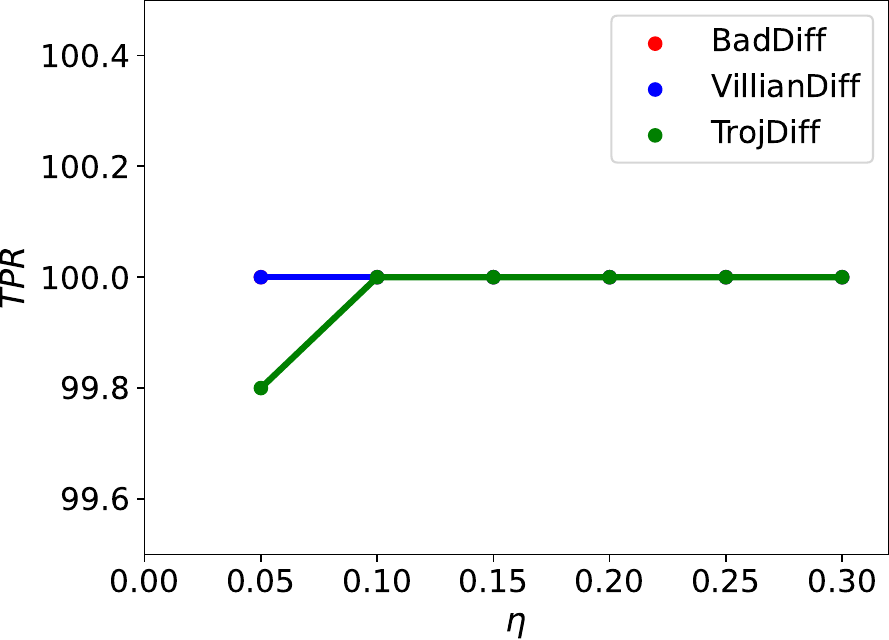}} 
    \subfigure[TNR]
{\includegraphics[width=0.385\linewidth]{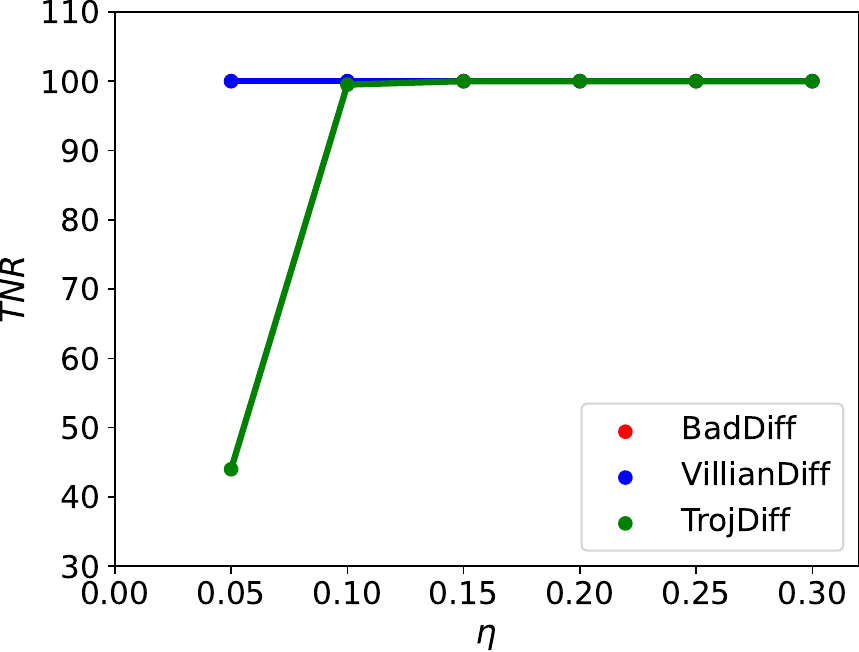}}
\caption{The performances of our proposed input detection method against the adaptive attacks.}
    \label{fig:input}
\end{figure}

\begin{table*}[h]
    \centering
    \caption{The generated images of benign noise with varied $\eta$.}
    \begin{tabular}{c| c| c| c|c|c|c}
    \toprule
        $\eta$&0.05&0.1&0.15&0.2&0.25&0.3\\
        \midrule
         BadDiffusion& \includegraphics[width=2cm,height=2cm]{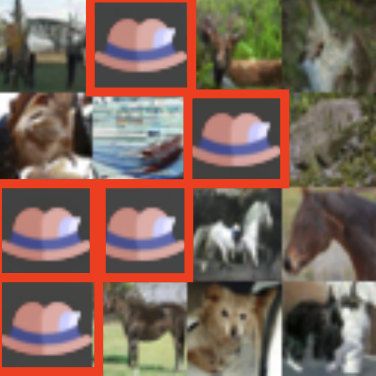}& \includegraphics[width=2cm,height=2cm]{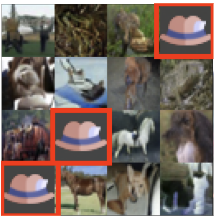}& \includegraphics[width=2cm,height=2cm]{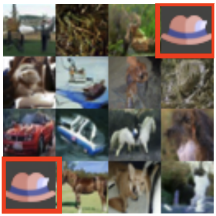}& \includegraphics[width=2cm,height=2cm]{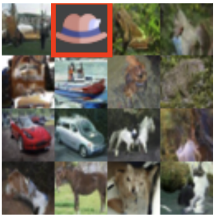}& \includegraphics[width=2cm,height=2cm]{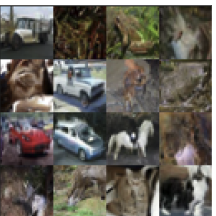} & \includegraphics[width=2cm,height=2cm]{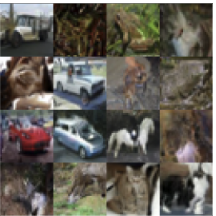}\\
        \midrule
         TrojDiff& \includegraphics[width=2cm,height=2cm]{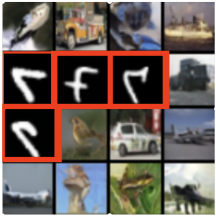}& \includegraphics[width=2cm,height=2cm]{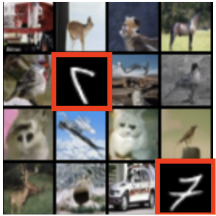}& \includegraphics[width=2cm,height=2cm]{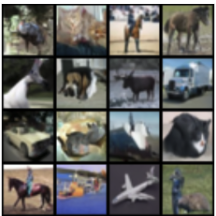}& \includegraphics[width=2cm,height=2cm]{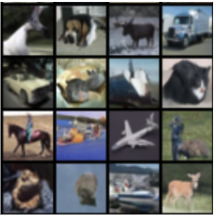}& \includegraphics[width=2cm,height=2cm]{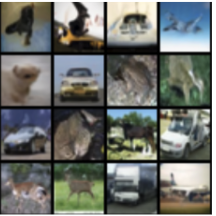}&  \includegraphics[width=2cm,height=2cm]{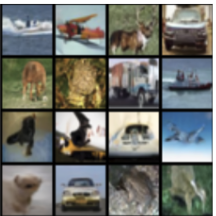}\\
         \midrule
         VillianDiffusion& \includegraphics[width=2cm,height=2cm]{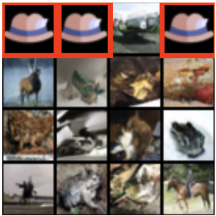}& \includegraphics[width=2cm,height=2cm]{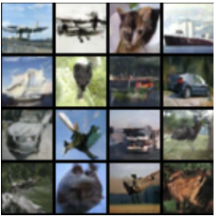}& \includegraphics[width=2cm,height=2cm]{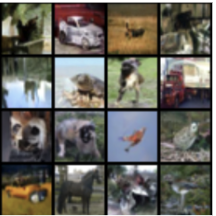}& \includegraphics[width=2cm,height=2cm]{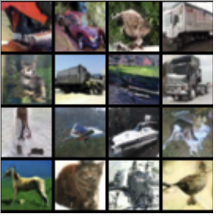}& \includegraphics[width=2cm,height=2cm]{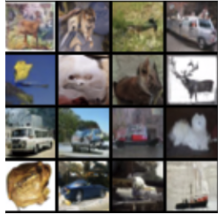}&  \includegraphics[width=2cm,height=2cm]{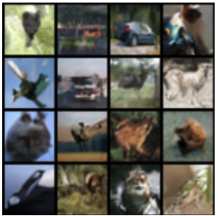}\\
    \toprule
    \end{tabular}
    \label{tab:clean_sample}
\end{table*}

\end{document}